\UseRawInputEncoding

\documentclass[conference]{IEEEtran}

\let\labelindent\relax
\usepackage{enumitem}
\usepackage{tikz}
\usepackage{xspace}
\usepackage{amsmath}
\usepackage{amsfonts}
\usepackage{bbm}
\usepackage{cite}
\usepackage{booktabs}
\usepackage{nameref}
\usepackage{multicol}
\usepackage{bm}
\usepackage{calc}
\usepackage{tabularx}
\usepackage{moreenum}
\usepackage{adjustbox}
\usepackage{flushend}
\definecolor{ref}{HTML}{116D6E}
\definecolor{cite}{HTML}{E55807}
\definecolor{takeaways}{HTML}{F5F5F5}

\DeclareMathAlphabet{\mathbf}{OMS}{cmsy}{m}{n}
\usepackage{hyperref}
\hypersetup{                    
  colorlinks,
  linkcolor={green!80!black},
  citecolor={red!70!black},
  urlcolor={blue!70!black}
}
\usepackage{xurl}
\usepackage{amssymb}
\usepackage{authblk}
\usepackage{xcolor} 
\usepackage{ragged2e}
\usepackage{colortbl}
\usepackage{float}
\usepackage{enumitem}
\usepackage{amsthm}

\usepackage{multirow}
\usepackage{pifont}
\usepackage[most]{tcolorbox}

\tcbuselibrary{breakable}
\usepackage{caption}
\makeatletter
\def\cmidrulewidth{\@cmidrulewidth}
\def\setcmidrulewidth#1{\gdef\@cmidrulewidth{#1}}
\makeatother
\usepackage{graphicx}  
\usepackage{subcaption} 
\usepackage{algorithm}
\usepackage{algpseudocode}

\allowdisplaybreaks

\begin{document}
\pagestyle{plain}

\renewcommand{\paragraph}[1]{\vspace*{6pt}\noindent\textbf{#1}\;}

\newcommand{\com}[2]{\noindent\textcolor{green}{\textbf{#1: }}\textcolor{blue}{#2}\xspace}
\newcommand{\py}[1]{\com{py}{{#1}}}
\newcommand{\says}[2]{\noindent\textcolor{orange}{\textbf{#1: }}\textcolor{purple}{#2}\xspace}
\newcommand{\tw}[1]{\says{tw}{#1}}

\renewcommand{\sectionautorefname}{Section}
\renewcommand{\subsectionautorefname}{Section}
\renewcommand{\subsubsectionautorefname}{Section}

\newcommand{\myPr}[1]{\ensuremath{\mathsf{Pr}\left[#1\right]}\xspace}
\renewcommand{\Pr}[1]{\ensuremath{\mathsf{Pr}\left[#1\right]}\xspace}

\newcommand{\fullsysname}{Latent Variable Defense\xspace}

\newcommand{\sysname}{\ensuremath{\mathsf{LVD}}\xspace}

\usetikzlibrary{shadings}
\newcounter{example}
\colorlet{colexam}{red!75!black}
\tcbset{
  base/.style={
    empty,
    frame engine=path,
    colframe=yellow!10,
    sharp corners,
    title={Takeaway},
    attach boxed title to top left={yshift*=-\tcboxedtitleheight},
    boxed title style={size=minimal, top=4pt, left=4pt},
    coltitle=colexam,fonttitle=\large\bfseries\sffamily,
  }
}

\newtcolorbox{takeaway}{%
  base,
  boxed title style={overlay={
    \draw[colexam,line width=3pt,] (frame.north west)--(frame.north east);
  }},
  colback=colexam,
  overlay unbroken={
    \draw[colexam] ([yshift=-1.5pt]title.north east)--([xshift=-0.5pt, yshift=-1.5pt]title.north-|frame.east);
  },
}

\definecolor{darkred}{rgb}{0.6, 0, 0}

\date{}

\title{Towards Understanding Unsafe Video Generation}

\author[$\dag$]{Yan Pang}
\author[$\S$]{Aiping Xiong}
\author[$\ddag$]{Yang Zhang}
\author[$\dag$]{Tianhao Wang}
\affil[ ]{
        $\dag$University of Virginia \quad $\S$ Penn State University \quad $\ddag$CISPA Helmholtz Center for Information Security
}
\affil[ ]{\{yanpang, tianhao\}@virginia.edu, axx29@psu.edu, zhang@cispa.de}

\maketitle
\begin{abstract}
Video generation models (VGMs) have demonstrated the capability to synthesize high-quality output. It is important to understand their potential to produce unsafe content, such as violent or terrifying videos. In this work, we provide a comprehensive understanding of unsafe video generation.

First, to confirm the possibility that these models could indeed generate unsafe videos, we choose unsafe content generation prompts collected from 4chan and Lexica, and three open-source SOTA VGMs to generate unsafe videos.
After filtering out duplicates and poorly generated content, we created an initial set of $2112$ unsafe videos from an original pool of $5607$ videos. Through clustering and thematic coding analysis of these generated videos, we identify $5$ unsafe video categories: \textit{Distorted/Weird}, \textit{Terrifying}, \textit{Pornographic}, \textit{Violent/Bloody}, and \textit{Political}. With IRB approval, we then recruit online participants to help label the generated videos. Based on the annotations submitted by $403$ participants, we identified $937$ unsafe videos from the initial video set. With the labeled information and the corresponding prompts, we created the first dataset of unsafe videos generated by VGMs. 

We then study possible defense mechanisms to prevent the generation of unsafe videos. Existing defense methods in image generation focus on filtering either input prompt or output results. We propose a new approach called \fullsysname (\sysname), which works within the model’s internal sampling process. \sysname can achieve $0.90$ defense accuracy while reducing time and computing resources by $10\times$ when sampling a large number of unsafe prompts. Our experiment includes three open-source SOTA video diffusion models, each achieving accuracy rates of $0.99$, $0.92$, and $0.91$, respectively. Additionally, our method was tested with adversarial prompts and on image-to-video diffusion models, and achieved nearly $1.0$ accuracy on both settings. {Our method also shows its interoperability by improving the performance of other defenses when combined with them.} We will publish our constructed video dataset\footnote{\url{https://huggingface.co/datasets/pypy/unsafe_generated_video_dataset}} and code\footnote{\url{https://github.com/py85252876/UVD}}.

\end{abstract}

\begin{justify}
\textcolor{darkred}{\textbf{Warning:}} \textcolor{red}{
This paper evaluates unsafe videos of generative AI and contains content that is sexual, offensive, etc.
}
\end{justify}

\section{Introduction}

Recently, video generation models (VGMs)~\cite{yuan2024magictime, blattmann2023stable, chen2024videocrafter2, guo2023animatediff, zhang2023i2vgen, zhang2023show, ho2022imagen, ho2022video} have improved significantly and can generate coherent and high-quality videos encompassing a wide variety of themes. 
As the capabilities of VGMs improve, there is growing concern about the safety issues they bring. For example, the ``Executive Order on the Safe, Secure, and Trustworthy Development and Use of Artificial Intelligence''\footnote{\url{https://www.whitehouse.gov/briefing-room/presidential-actions/2023/10/30/executive-order-on-the-safe-secure-and-trustworthy-development-and-use-of-artificial-intelligence/}} from the White House emphasizes ``{testing and safeguards against discriminatory, misleading, inflammatory, unsafe, or deceptive outputs}'' in Section $10$; In April 2024, NIST released the AI Risk Management Framework, which explicitly tries to understand the ability of generative models to synthesize unsafe content. We noticed on websites like Civitai\footnote{\url{https://civitai.com/}} that users can share their prompts and generated content. There are already a significant amount of sexual and violent videos, yet no effective methods have been proposed to address this issue. 

To examine the capacity of VGMs to generate unsafe videos, we use prompts collected from the 4chan and Lexica websites~\cite{schramowski2023safe, qu2023unsafe}; These datasets were originally used to guide text-to-image models in generating unsafe images. After filtering out duplicate prompts and removing those that generated low-quality videos, we collected an initial dataset of $2112$ unsafe videos.

To verify these generative videos are indeed unsafe and cause unpleasant feelings, with IRB approval, we recruited online study participants to rate those videos.  Specifically, we first clustered unsafe videos and did the thematic coding analysis~\cite{braun2006using}. Through two rounds of discussions, we identified five main categories of unsafe videos: \textit{Distorted/Weird}, \textit{Terrifying}, \textit{Pornographic}, \textit{Violent/Bloody}, and \textit{Political}. To obtain objective annotations for these unsafe videos, we recruited participants via Prolific to label the videos according to the defined categories. We initially recruited $600$ participants. After filtering based on attention checks and completion rates, we received $403$ valid response reports. Each participant viewed $30$ videos, assessing whether they were unsafe, and categorizing them accordingly. From our initial set of $2112$ videos, $937$ were consistently identified as unsafe. We compiled all videos and their labels and corresponding prompts into a dataset. This is the first dataset of unsafe videos generated by VGMs.

Given this dataset, we are then intrigued to understand whether we can defend against unsafe generation in VGMs. That is, can we ensure VGMs will never generate unsafe content?
Note that this is related to defense against deepfakes, but deepfakes are primarily focusing on facial videos~\cite{guera2018deepfake,wodajo2023deepfake,dang2020detection,gandhi2020adversarial,he2021forgerynet,wang2020fakespotter}, and our focus is VGMs in general (various types, not just facial videos). 
There are existing solutions~\cite{qu2023unsafe,li2024safegen,schramowski2023safe,li2022nsfw,rando2022red,gandikota2023erasing,brack2024sega, kumari2023ablating, kim2023towards, lyu2024one, gandikota2024unified} for image generative models that addressed the unsafe generation problems. 
For example, Schramowski et al.\cite{schramowski2023safe} designed a safety guidance strategy that uses pre-defined safety prompts to redirect potentially harmful prompts. Then, Gandikota et al.\cite{gandikota2023erasing} suggested removing harmful concepts from the model's understanding by fine-tuning the entire model. Li et al.\cite{li2024safegen} noted the text-dependency of previous methods and proposed removing unsafe visual representations to protect generated content. Qu et al.\cite{qu2023unsafe} implemented a detection model to evaluate the output content without interfering with the generation process. However, detecting unsafe content in videos is much more challenging than in images because videos contain more information (both spatial and temporal) and require significantly more computational resources to generate. Therefore, we need to rethink methods for detecting unsafe content in the video domain.

\begin{figure}[t]
    \centering
    \resizebox{0.47\textwidth}{!}{
    \includegraphics{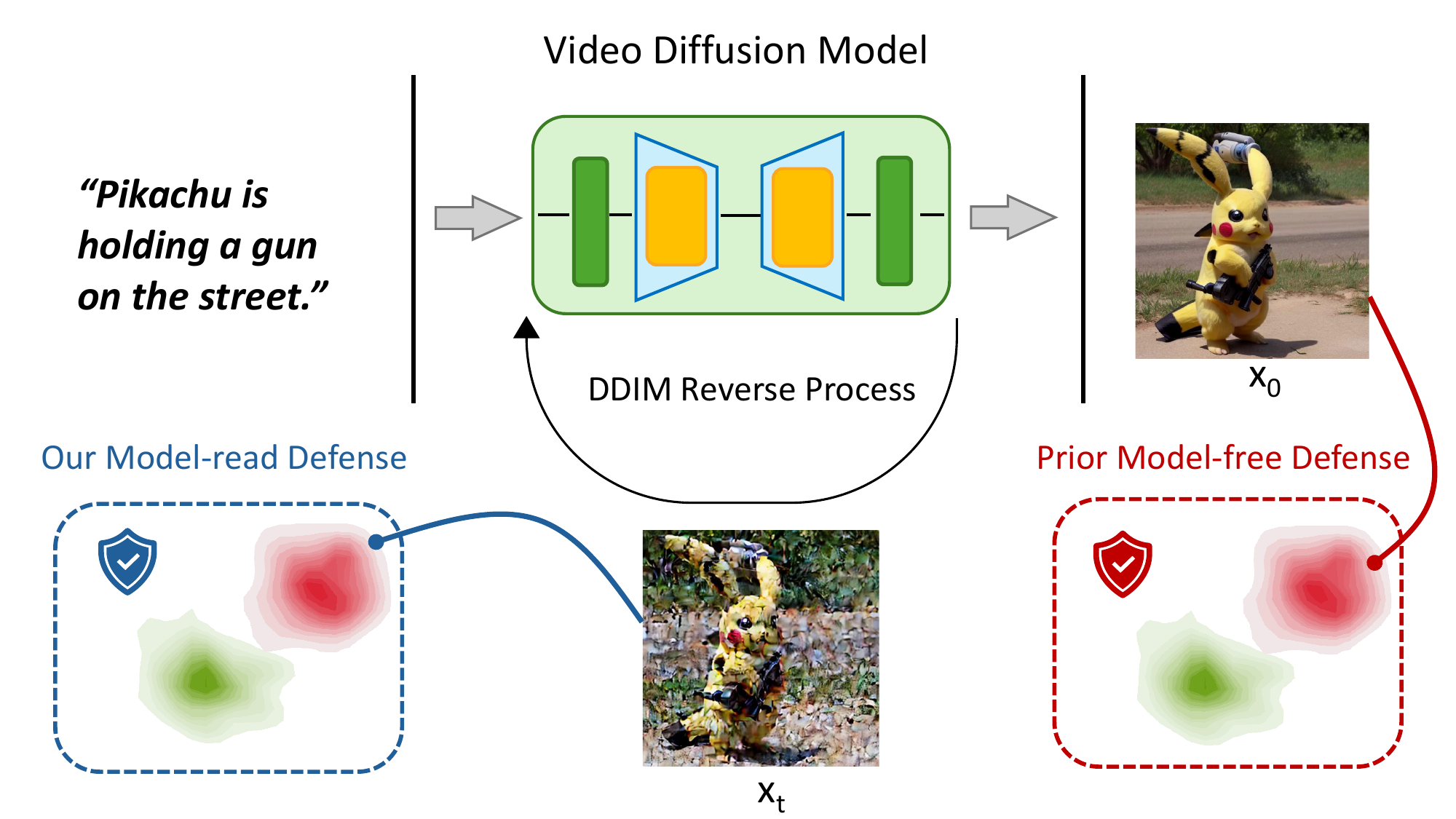}}
    \caption{Unlike previous model-free defense methods for image diffusion models, we proposed utilizing the DDIM sampler's deterministic characteristics and using the intermediate denoising outputs to assess whether the generated video is unsafe. See detailed description in~\autoref{sec:Methodology}.}
    \label{fig:pipeline}
\end{figure}

As detailed in~\autoref{sec:Methodology}, we categorize existing solutions for image generative models as either model-free (no need to look into model internals) or model-write (need to modify model parameters or settings) approaches. These two types of methods can only employ the final results of the diffusion process, limit further defensive actions, or require significant computational resources to update model parameters. Nevertheless, we propose a middle-ground, {\it model-read} approach that leverages ``read access'' of the diffusion model and detects unsafe content within the diffusion process. The {\it model-read} method is less rigid than the model-free approach, which only detects the model's final output. It also avoids the extensive time and computational resources needed for the model-write approach. We call our method \fullsysname (\sysname). \sysname leverages the intuition that generative models, such as VAE~\cite{kingma2013auto} and diffusion models~\cite{ho2020denoising}, are trained to learn latent space representation, and in these representations, samples that are close in the latent space result in similar generated content~\cite{kingma2013auto}. As shown in~\autoref{fig:pipeline}, we establish classifiers to analyze the intermediate results of the diffusion process, which is quite different from the existing model-free methods. When we suspect the result will be unsafe, we can terminate the diffusion process early, saving significant computation resources. All Experimental results on three SOTA VGMs show that compared to working with the final result, our approach can save up to $10\times$ computation time, while achieving comparable detection accuracy (around $92\%$). 




\paragraph{Contributions:} We make the following contributions.
\begin{itemize}[leftmargin=*]
       
    \item 
    We demonstrated that VGMs have a strong capability to synthesize unsafe content. 
    \item We construct an unsafe generated video dataset from the SOTA open-sourced VGMs. Unsafe categories were generated using data-driven methods. Data annotations were completed by $403$ participants recruited through Prolific.
    \item Based on our unsafe video dataset, we proposed \sysname to mitigate generating unsafe videos. \sysname uses DDIM characteristics to detect unsafe content during inference time.
    \item We tested our defense on three open-source SOTA VGMs, conducting comprehensive experiments to assess the defense performance under different parameter settings. Results showed our defense can achieve nearly $100\%$ accuracy on all three models. We further assessed our defense's robustness, generalization ability, and interoperability to show the effectiveness of our method.

\end{itemize}

\paragraph{Roadmap.} The remaining of the paper is organized as follows. In~\autoref{sec:background}, we introduce the basic concept of diffusion models and mention defenses against deepfake attacks. Then, in~\autoref{sec:UVG}, we discuss how we built our unsafe video dataset. Based on this dataset, we design our defense mechanism in~\autoref{sec:Methodology}. ~\autoref{sec:safety_assessment} provides details about how we examined our defense mechanism. To further test it, we conducted ablation studies in~\autoref{sec:ablation}. ~\autoref{sec:Related_work} summarizes related work, and we discuss the limitations and conclude in~\autoref{sec:Conclusion}.

\section{Background} \label{sec:background}

\subsection{Diffusion Models} \label{sec:bk_dm}

Diffusion models~\cite{ho2020denoising, nichol2021improved} are state-of-the-art generative models that have been used in various modalities, such as image~\cite{rombach2022highresolution, saharia2022photorealistic}, audio~\cite{liu2023audioldm}, and video generation~\cite{blattmann2023stable,chen2024videocrafter2, zhang2023i2vgen}. The fundamental concept behind diffusion models consists of two phases: the diffusion process and the denoising process. The diffusion process, also called the forward process, iteratively adds noise from a standard normal distribution. 

The noise schedule $\{{\alpha_t}\}^{T}_{t=1}$ is set to control the magnitude of noise added in the diffusion process. Utilizing a reparameterization trick, we can express the noise sample $x_t$ at any step $t$ in the forward pass, given original data $x_0$, as:
\begin{equation} \label{equ:x_t}
    x_t = \sqrt{\bar{\alpha}_t}x_0 + \sqrt{1-\bar{\alpha}_t}\epsilon_t
\end{equation}
where $\bar{\alpha}_t = \prod_{i=1}^{t} \alpha_i$.

In contrast, the denoising process aims to remove noise from a noisy image $x_T$, where $x_T \sim \mathcal{N}(0,1)$ and ultimately denoise to a clean image $x_0$. A neural network (e.g., U-Net) $\epsilon_{\theta}$ is usually trained to predict the noise that needs to be removed at step $t$. The loss function for training the denoising network can be represented as:
\begin{align}
    L_t(\theta) = \mathbb{E}_{x_0,\epsilon_t}\left[ \| \epsilon_t- \epsilon_{\theta}(\sqrt{\bar{\alpha}_t}x_0 + \sqrt{1-\bar{\alpha}_t}\epsilon_t,t)\|^2_2\right]
\end{align}

\subsection{Video Diffusion Models}

To apply diffusion models~\cite{ho2020denoising,song2020denoising} to video generation, these models need to understand spatial information and maintain consistency and coherence in the temporal dimension. Unlike traditional diffusion models~\cite{ho2020denoising,song2020denoising}, which work mostly in 2D, VGMs must incorporate temporal layers to learn the motion logic of the object over time~\cite{ho2022video}. With the new requirement for generative objectives, different approaches to training VGMs exist. Based on training strategies, we categorize them into \textit{training from scratch}, \textit{fine-tuning on video data}, and \textit{training-free models}. 

Among the three approaches, fine-tuning stands out as an efficient strategy and demonstrates superior performance and becomes the focus of this study (and we will elaborate on the other two approaches in~\autoref{sec:moretraining}). Most models adopt this approach, as it balances synthesis quality and the consumption of training resources. This approach requires selecting an appropriate pre-trained image generator as the backbone. And currently, existing works~\cite{zhang2023i2vgen,blattmann2023stable,chen2024videocrafter2,yuan2024magictime} usually select stable diffusion~\cite{rombach2022highresolution} as the backbone. 
After adding temporal convolution and attention layers to ensure consistency across multiple frames, the model can be trained directly using video data. During training, researchers need to freeze the parameters of the spatial layers and focus on training the inserted temporal layers to learn the motion logic of objects. Once the model understands temporal dynamics, the generation quality can be further improved using a cascade approach. This involves dividing the generation process into a base stage and a refine stage~\cite{ho2022imagen}. In the base stage, a large amount of low-resolution video data helps the model understand the generation objective and produce a low-quality video. In the refining stage, a small number of high-resolution video data is used to train the model. This stage enhances the quality of the video generated in the base stage, ultimately achieving good generation results.



\paragraph{Deepfakes.}
One kind of risky generated video that draws extensive attention is human facial videos, a.k.a. deepfakes.
Deepfake attacks are used to primarily target facial images~\cite{wang2020cnngenerated,zhang2019detecting,zhu2023genimage, cozzolino2023synthetic} and videos~\cite{guera2018deepfake,wodajo2023deepfake,dang2020detection,gandhi2020adversarial,he2021forgerynet,wang2020fakespotter}. However, with the development of VGMs~\cite{zhang2023i2vgen,guo2023animatediff,yuan2024magictime,chen2024videocrafter2}, it is now possible to create high-resolution realistic scenes. We differentiate ourselves from Deepfakes as the generated topic is no longer limited to facial images.

\subsection{Threat Model}

Our study assumes a simple threat model that involves only two parties: malicious users and model owners. 
The goal of the malicious user is to use the video generation model to synthesize unsafe videos. These users have unsafe prompts and tried to feed those prompts to the model. Moreover, malicious users potentially have the capability to use the optimization method to build adversarial prompts. They can access the model output but cannot access the internal values.

On the other hand, model owners can access the model's internal parameters and get intermediate output. Their goal is to design an effective method to protect their model from being abused by malicious users while not affecting its generating ability on normal prompts.


\section{Generate Unsafe Videos} \label{sec:UVG}

We first explore the feasibility of video generative models (VGMs) to synthesize unsafe videos. 
With positive results, we then recruited participants to identify and label unsafe videos, constructing an unsafe video dataset.

\subsection{Unsafe Prompt Collection}

The prompts we choose to generate unsafe content are from two unsafe prompt datasets: the dataset from Qu et al.~\cite{qu2023unsafe} and the I2P (Inappropriate Image Prompts) dataset~\cite{schramowski2023safe}. These two datasets contain many unsafe prompts, and all have been tested on the text-to-image generation models. Since most current VGMs~\cite{guo2023animatediff,blattmann2023stable,yuan2024magictime,chen2024videocrafter2,zhang2023i2vgen} still use the pre-trained T2I model~\cite{rombach2022highresolution} as their backbone and have the same spatial understanding, we first used those datasets to explore the capabilities of VGMs. Based on the content quality, we then selected prompts to build our experimental dataset.

Specifically, Qu et al.~\cite{qu2023unsafe} collect their unsafe prompts from 4chan\footnote{\url{https://www.4chan.org/index.php}} and Lexica.\footnote{\url{https://lexica.art/}} 4chan is an anonymous platform where people post unsafe content, including sexual, violent, hateful, etc. After preprocessing the raw data from 4chan, they collected $500$ prompts that could generate high-quality images from 4chan. Different from the 4chan website, Lexica is a website that provides prompts directly. Thus, they used $66$ keywords related to the unsafe categories to query the Lexica website. After data cleaning, they collected $404$ prompts from Lexica.

The I2P dataset is collected by Schramowski et al.~\cite{schramowski2023safe}, and its prompts are also from the Lexica website. The authors also used keywords related to their seven unsafe categories to query Lexica. They designed $26$ keywords and collected $250$ text prompts for each word. Because some prompts can have multiple keywords simultaneously, they finally collected $4703$ prompts in their datasets after removing the duplicate prompts.

\begin{figure*}[!t]
    \centering
    \resizebox{0.99\textwidth}{!}{
    \includegraphics{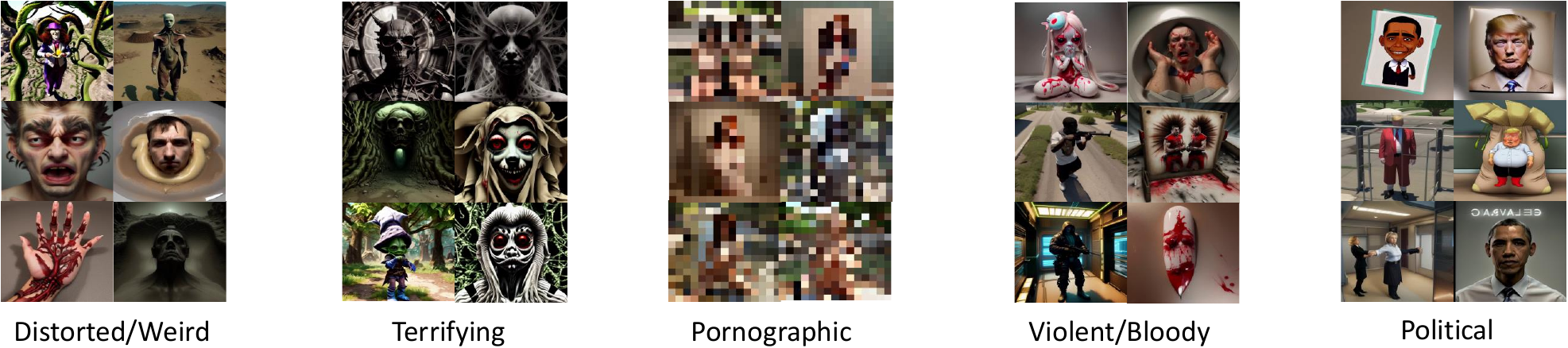}}
    \caption{Based on our thematic coding analysis, we identified five categories of unsafe videos from the generated videos. For each category, we selected the first frame of the representative videos to illustrate our findings. For the \textit{Pornographic} videos, we add masks to cover the explicit sexual content.}
    \label{fig:generated_img}
\end{figure*}

\subsection{Theme Summary}

In our work, we first combined all the unsafe prompts, then removed the duplicate prompts from two datasets, and used those prompts to query MagicTime~\cite{yuan2024magictime} to generate the corresponding videos. Because our work is on the video domain and the model differs from the previous work~\cite{gandikota2023erasing,saharia2022photorealistic,qu2023unsafe,schramowski2023safe}, we cannot directly use the unsafe categories defined in their paper.

Similar to the previous work~\cite{qu2023unsafe}, we want to use a data-driven approach to identify the scope of unsafe images. First, we used the $k$-means method~\cite{sinaga2020unsupervised} to categorize the generated videos into several clusters. Then, we selected the most optimized cluster number and did the thematic coding analysis to define the unsafe category's name.

Because image feature extractors are much more powerful than video feature extractors due to the amount of training data, we mainly focus on the semantic information for each video. Thus, we selected the CLIP model (CLIP-ViT-L-14) as our feature extractor and processed all the frames for each video. We calculated the mean of all frame feature vectors in each video to get the feature vector representing the video. Then, we used the $k$-means clustering algorithm to get the video feature clusters and set the possible cluster number from $2$ to $30$. According to the elbow method~\cite{syakur2018integration}, we found that when we set the cluster number to $23$, the generated videos would have the best cluster performance.

The next step is to identify unsafe video categories. Since no unsafe video detector currently exists, we cannot remove unsafe videos before performing clustering. Some video clusters consist of normal videos that require manual inspection. We applied thematic coding analysis~\cite{braun2006using} in our work, which is usually used in social science, to conclude the theme by qualitatively analyzing data. To get better results, we collected $10$ videos from each cluster, and three authors of our work wrote the text description for each cluster. This is the initial version of our code book, and then we calculate the initial Krippendorff's alpha~\cite{hayes2007answering} is $0.39$, and Fleiss' kappa~\cite{fleiss1971measuring} is $0.56$. Then, we discussed the code and tried to refine our code book. We built an overall text description for each cluster and changed some of our initial code based on that. For our second code book, Krippendorff's alpha achieves $0.83$, and the score of Fleiss' Kappa is $0.94$. Both of these scores represent that we had an agreement for almost every cluster. The final step is to group the clusters, as shown in~\hyperref[appendix:Data_Collect]{Appendix~\ref*{appendix:Data_Collect}}. Although some videos are from different clusters, they can be the same type of unsafe videos. After removing the harmless video cluster, we conclude five unsafe categories: \textit{Distorted/Weird}, \textit{Terrifying}, \textit{Pornographic}, \textit{Violent/Bloody}, and \textit{Political}. We show the videos represented for each category in~\autoref{fig:generated_img}. In our work, we use these five unsafe categories to classify unsafe videos and design our defense.

\subsection{Data Collection} \label{sec:data_collection}

We designed an online survey based on the five categories of unsafe videos identified through thematic coding analysis~\cite{braun2006using}. Our survey contained $2,112$ unsafe videos produced by MagicTime~\cite{yuan2024magictime}. Unlike previous studies that used authors to label inappropriate content, we aimed to reduce subjective bias and gather more authentic and neutral data. Initially, we obtained approval from the institute's IRB protocol and subsequently created the survey using Qualtrics website. Participants for annotating unsafe videos were recruited through Prolific platform, and compensation was provided. Their demographic data will be protected during this process.

Given the sensitive nature of our survey, we included a disclaimer on the first page of Qualtrics and the Prolific homepage, stating that the content contains unsafe information. Participants needed to be over $18$ years old and were informed that they could withdraw from the survey at any time if they felt uncomfortable without facing any penalties. Each participant was assigned $30$ videos and one additional attention check. If participants considered a video to be unsafe, they were prompted to categorize it into one of five categories: \textit{Distorted/Weird}, \textit{Terrifying}, \textit{Pornographic}, \textit{Violent/Bloody}, and \textit{Political}. An ``Other'' option was also provided for any videos that did not fit these categories. Participants were compensated at a rate of $\$10.15/\text{hr}$.

We first conducted a pilot study to validate the design and logic of our Qualtrics survey. We recruited $20$ participants to test our questionnaire. The goal of the pilot study was to verify the appropriateness of our time settings, the number of questions, and the attention checks. After examining the responses submitted by each participant, we found that there were no necessary modifications needed for the questions or the attention check. Accordingly, we decided to recruit $600$ participants for the main survey study. These $600$ participants for the main survey did not overlap with those who participated in the pilot study.

From the initial pool of $600$ participants, $197$ individuals failed the attention check and were subsequently excluded from the study. For these participants, we still provided $10\%$ of their compensation through bonus payments. As a result, a total of $403$ participants successfully completed the labeling task for $30$ generated videos. We first categorized all videos deemed ``unsafe video'' according to participants' labels. Specifically, every video was labeled by at least two participants in our survey. Next, we performed data cleaning by integrating our own assessments of the ``unsafe video'' videos. For example, if a video was labeled as ``unsafe'' and categorized as \textit{Pornographic} (label $4$), it was retained in the \textit{Pornographic} category only if more than half of the participants who marked it as ``unsafe video'' also identified it as \textit{Pornographic}. After data cleaning, the videos were categorized based on participant labels and our assessments. We got $590$ videos as \textit{Distorted/Weird}, $579$ as \textit{Terrifying}, $445$ as \textit{Pornographic}, $204$ as \textit{Violent/Bloody}, and $39$ as \textit{Political}.

We allowed participants to suggest new unsafe categories during our online study. However, because there was not a sufficient number of participants who reached a consensus on any additional category, and because some of the suggested categories were similar to our defined ones (e.g., sexually explicit), we maintained the five unsafe categories generated through thematic coding analysis.

\section{Defense Methodology} \label{sec:Methodology}

Several studies have discussed potential safety issues within image generation models, and proposed defense methods. We group these existing defense methods into two clusters: model-write defense~\cite{gandikota2023erasing,li2024safegen,schramowski2023safe, brack2024sega, kumari2023ablating, kim2023towards, lyu2024one, gandikota2024unified} that modify part of the model weight or generation process, and model-free defense~\cite{qu2023unsafe,li2022nsfw,rando2022red} that based on input and/or output filtering and does not require access to the model.  

Model-write defenses change or update the model's generation process or parameters, which might affect the model generation quality, and usually need to fine-tune the model, which takes time~\cite{gandikota2023erasing,li2024safegen, kumari2023ablating, kim2023towards, lyu2024one, gandikota2024unified}. On the other hand, model-free defense trains classifiers/detectors that predict whether the input prompt and/or output result is harmless. Input prompt filtering is vulnerable to adversarial prompts and jailbreak attacks (described briefly in~\autoref{sec:jailbreak}), while output filtering still needs model owners to first generate the results and also takes time (generating videos takes tens of times longer than generating images).
We provide more details of existing defenses for image diffusion models in~\autoref{sec:existing_defense_for_image}.



\subsection{Overview of Our Method} \label{sec:overview_method}
Given the drawbacks of model-write and model-free approaches, we propose a {\it model-read} approach that sits between model-write and model-free defenses.  At a high level, we only need ``read access'' of the diffusion model and detect unsafe content within the diffusion process.  It shares similarities with the model-free approach in that both train detectors, but it generalizes the model-free approach by considering both input filtration and output filtration as extreme cases. Model-read defense is more robust compared to input filtration and more efficient compared to output filtration (because it does not require completing the entire generation process). Note that our model-read defense can also potentially collaborate with the other two approaches (as discussed in~\autoref{sec:inter}), providing a more comprehensive defense.

Our solution relies on the insight that generative models, like VAE~\cite{kingma2013auto} and diffusion models~\cite{ho2020denoising}, are designed to learn latent space representations. We conjecture that the same type of unsafe videos are generated by latent variables close to each other in the latent space.







\subsection{\fullsysname} 
In this section, we introduce our defense called \fullsysname (\sysname). The foundation for our method is the DDIM sampler~\cite{song2020denoising} used in modern diffusion models for video/image generation, which can significantly enhance inference speed. 

\paragraph{DDIM Foundation.} Compared to the traditional DDPM's Markovian sampling process~\cite{ho2020denoising}, DDIM is non-Markovian and deterministic. Different video diffusion models may vary in structure due to distinct design choices~\cite{chen2024videocrafter2,yuan2024magictime,guo2023animatediff}. However, to efficiently generate samples, these models move the diffusion process to the latent space. 
The reverse process in these models can be represented as:
\begin{align}
    z_{t-1} = & \sqrt{\bar{\alpha}_{t-1}}\left( \frac{z_t-\sqrt{1-\bar{\alpha}_t} \epsilon_{\theta}(z_t,t)}{\sqrt{\bar{\alpha}_t}}\right) \notag \\
    & + \sqrt{1-\bar{\alpha}_{t-1}-\sigma_t^2}\cdot\epsilon_{\theta}(z_t,t) \notag + \sigma_t\epsilon_t
\end{align}
where $\sigma_t$ denotes the hyper-parameter that controls the level of randomness in the forward process~\cite{song2020denoising}, and other symbols share the same notations as in~\autoref{sec:bk_dm}. When the $\sigma_t = 0$ for all $t$, the whole reverse/forward process is a deterministic trajectory. Song et al.~\cite{song2020denoising} prove that when both reverse and forward processes are fixed, model sampling steps can be accelerated by defining the reverse process on a subset of the original $T$ steps and still get the high-quality output. For a video/image diffusion model, the inference steps can be accelerated to only $k$ steps $\tau = \{\tau_1,\tau_2,...,\tau_k\}$ (these $k$ steps evenly partition the original $T$ steps).  The accelerated DDIM sampler equation can be represented as:
\begin{equation}
    \Pr{z_{\tau_{i}}|z_{\tau_{i+1}},z_0} = \frac{\Pr{z_{\tau_{i+1}}|z_{\tau_{i}},z_0}\Pr{z_{\tau_{i}}|z_0}}{\Pr{z_{\tau_{i+1}}|z_0}}
\end{equation} 
where $\tau_{i}<\tau_{i+1}-1$ and $\tau_k\le T$; $z$ refers to the latent variable. For each latent variable $z_{\tau_i}$ in the reverse process, it does not have randomness and is controllable. Given the specific latent variable $z_{\tau_{i}}$, the final output $z_0$ is also constant. Therefore, we want to use the DDIM deterministic denoising trajectory to design our defense method. 

Unlike the previous model-free methods~\cite{qu2023unsafe,li2022nsfw,rando2022red}, which rigidly used the final synthesis images as input for detection, we want to use the intermediate latent variables from $z_{\tau_k}$ to $z_{\tau_1}$ to train our detection model $M$. This approach will help build a more agile and integrable defense mechanism.


\paragraph{Defense Algorithm.} Assume the video generation model sets the number of inference steps to $k$ out of $T$ (according to the existing literature~\cite{song2020denoising}, setting the number of denoising steps to $k=50$ achieves nearly the same quality as $T=1000$ steps, saving time and computing resources; we also set the denoising step to $50$ in our experiment), our defense mechanism involves $k$ detection models $M_{1},{M}_2,...,{M}_k$. We use the latent variable at $i$-th step to train each ${M}_i$. It is important to note that we do not directly feed the $z_{\tau_t}$ at $t$-th step into our detection model. According to~\autoref{equ:x_t}, we can calculated denoised sample $z^{0}_{\tau_t}$ at $t$-th step and use it as input for detection model. The detection results for $i$-th model can be represented as $s_i = {M}_i(z^{0}_{\tau_i})$. For each data point, our mechanism can obtain a vector of scores $s_1,..,s_k$ that helps us determine whether the current video is an unsafe video.  


We give the details of \sysname in~\autoref{alg:defense}.  Specifically, it leverages detection models to determine at each step whether the intermediate result is unsafe and makes the final decision based on cumulative scores. We introduce two hyperparameters, $\lambda$ and $\eta$. $\eta$ improves \sysname's efficiency by considering only the first $\eta<k$ steps, and $\lambda$ controls the detection threshold.



\begin{algorithm}[!t]
    \caption{\fullsysname (\sysname)}
    \label{alg:defense}
    \begin{algorithmic}[1]
    \Require Input prompt $p$, $k$ detection models ${M}_1,\ldots,M_k$, a set of sampling steps $\tau_1$,\ldots,$\tau_k$, and defense parameters $\lambda$, and $\eta$. $D$ is the decoder in the video generation model. 
    \State Sampling the initial latent variable $z_{{\tau}_{k+1}} \sim \mathcal{N}(0,1)$ 
    \For{$i\gets k$ \textbf{to} $k-\eta+1$} \Comment{Perform $\eta$ steps.}
    \State $z_{{\tau}_i}$ $\leftarrow$ $\epsilon_{\theta}(z_{{\tau}_{i+1}},\tau_{i+1})$
    \State Get denoised $z^{0}_{{\tau}_i}$ from $z_{{\tau}_i}$ \Comment{By~\autoref{equ:x_t}}
    \State $s_i \leftarrow {M}_{i}(D(z^{0}_{{\tau}_i}))$ \Comment{$1$: unsafe, $0$: safe.}
    \EndFor
    
    \Ensure $\mathbbm{1} \left(\sum_{j=i}^\eta s_j \geq \lambda \cdot \eta\right)$\Comment{$1$: unsafe, $0$: safe.}
    \end{algorithmic}
\end{algorithm}

To further improve efficiency, \sysname can dynamically check the score vector for the generated sample. At the \(i\)-th step, the score for the generated video is given by $\text{score} = \sum_{j=i}^{k} {M}_{{j}}(z^{0}_{\tau_{j}})$. If the score is greater than or equal to $\lambda \cdot (k-i+1)$, \sysname can break and immediately classify the current video as unsafe. 



\subsection{Interoperability with Existing Defenses}
\label{sec:inter}
Since our method is applicable to all diffusion models, we will briefly demonstrate how it can be combined with existing defense methods. We will showcase this using both model-write~\cite{schramowski2023safe} and model-free methods~\cite{qu2023unsafe}.

\paragraph{Interoperate with Model-free Defense.} For model-free defense, integrating with our method is straightforward because the detection outputs differ. For example, in Qu et al.~\cite{qu2023unsafe}'s Unsafe Diffusion, with a diffusion model set to $k$ denoising steps, the combined defense pipeline can be represented as:
\begin{align}
     \mathcal{M}_p(G(p)) =& \gamma \cdot \mathbbm{1} \left( \sum_{i=1}^{k} {M}_{i}(z^{0}_{\tau_i}) \geq \lambda \cdot k \right) \notag \\  +& (1-\gamma)\left(\mathcal{M}_{u}(D(z^{0}_{\tau_k})\right)\notag
\end{align}
Here, $D$ represents the decoder part of the VGM, and $\gamma$ is a hyperparameter used to balance the combined system. We denote the multi-headed safety classifier from Unsafe Diffusion as $\mathcal{M}_{u}$. $\mathcal{M}_{u}$ takes the output sample from the decoder, $D(z^{0}_{\tau_k})$, as input for evaluation. 

\paragraph{Interoperate with Model-write Defense.}
For the model-write defense methods, besides updating the model parameters~\cite{li2024safegen,schramowski2023safe, kumari2023ablating, kim2023towards, lyu2024one, gandikota2024unified}, we can combine our approach with text-dependent methods~\cite{schramowski2023safe, brack2024sega}. In this section, we will briefly discuss how to integrate our method with SLD~\cite{schramowski2023safe}. The primary goal of SLD is to shift unsafe concepts during the inference step. In their original work, they use the difference between noise from the safety prompt $s$ and the input prompt $p$ to determine whether to guide the generation direction in the opposite direction. However, to keep the changes minimal, they set a warmup step $\delta$. This is because the noise difference is significant at the beginning of the inference process. In the early stages of inference, our \sysname can achieve high detection accuracy. Therefore, we can replace the calculation of noise difference $\mu(p,s)$ at each step to our \sysname. At $t$-th step, $\bar{\epsilon}_{\theta}(z_t,c_p,c_s) = $
\begin{equation}
    \begin{cases}
        \epsilon_{\theta}(z_t,c_p) + w(\epsilon_{\theta}(z_t,c_p) - \epsilon_{\theta}(z_t,c_s)), & \text{if } \mathcal{M}(z_t;\eta,\lambda) > \beta \\
        \epsilon_{\theta}(z_t,c_p) + w(\epsilon_{\theta}(z_t,c_p) - \epsilon_{\theta}(z_t)), & \text{otherwise}
    \label{eq:internal}
    \end{cases}
\end{equation}
where $c_p$ and $c_s$ are text embedding for $p$ and $s$, $\beta$ is the pre-defined threshold value, and $w$ is the guidance scale. In~\autoref{eq:internal}, when \sysname detects the generated content is unsafe, the VGM can redirect the generation process that is opposite to the safety concept.

\section{Evaluation} \label{sec:safety_assessment}

\subsection{Experiment Setup}

\paragraph{Data Preparation.}As mentioned in~\autoref{sec:data_collection}, we recruited $600$ participants via the Prolific platform to label $2,112$ videos generated by MagicTime~\cite{yuan2024magictime}. 
Moreover, since AnimateDiff~\cite{guo2023animatediff} and VideoCrafter~\cite{chen2024videocrafter2} also use Stable Diffusion~\cite{rombach2022highresolution} as their backbone (MagicTime and AnimateDiff used SD v1-5, and VideoCrafter used SD 2.1; their semantic-level understanding of unsafe prompts are similar). After carefully reviewing the label information from participants, we annotated the videos generated from the other two models. The number of videos in different unsafe groups generated by the various video diffusion models is presented in~\autoref{tab:unsafe_num}. In the evaluation process, we used $20\%$ unsafe videos to build an examination dataset to test our defense accuracy. We set $k = 50$ following existing literature~\cite{song2020denoising}, as $k = 50$ can already ensure high-quality generation while saving significant time.

\begin{table}[t]
    \belowrulesep=0pt
    \aboverulesep=0pt
    \Huge
    \centering
    \resizebox{0.48\textwidth}{!}{
    \begin{tabular}{c|ccccc|c}
    \toprule[2.8pt]
    \multirow{2}{*}{Model} & \multirow{2}{*}{\shortstack{Distorted\\ or Weird}} & \multirow{2}{*}{{Terrify}} & \multirow{2}{*}{{Porn}} & \multirow{2}{*}{\shortstack{Violent\\ or Bloody}} & \multirow{2}{*}{Political} & \multirow{2}{*}{Total} \\ 
    & & & & & & \\ 
    \midrule[0.5pt]
    MagicTime~\cite{yuan2024magictime} & $590$ & $579$ & $445$ & $204$ & $39$ & $937$\\ 
    VideoCrafter~\cite{chen2024videocrafter2} & $571$ & $564$ & $353$ & $197$ & $79$ & $931$\\
    AnimateDiff~\cite{guo2023animatediff} & $586$ & $577$ & $391$ & $204$ & $75$ & $945$ \\
    \bottomrule[2.8pt]
    \end{tabular}}
    \caption{Number of different categories of unsafe videos generated by different models.}
    \label{tab:unsafe_num}
\end{table}

\paragraph{Observation about Models.} Upon examining the different types of unsafe videos generated by these models, as presented in~\autoref{tab:unsafe_num}, we observed an intriguing phenomenon: the models exhibit variability in generating different categories of unsafe videos. Specifically, we found that MagicTime~\cite{yuan2024magictime} tends to produce higher-quality \textit{Pornographic} videos than the other two models. When the same prompts are fed into VideoCrafter~\cite{chen2024videocrafter2} and AnimateDiff~\cite{guo2023animatediff}, there is a certain likelihood of generating low-quality, chaotic videos or content lacking \textit{Pornographic} elements. Conversely, MagicTime is less accurate in handling political prompts, generating nearly half the number of political videos compared to the other two models. Noting these variations in the models' generation capabilities for different features, we selected these three current open-source state-of-the-art models to evaluate the effectiveness of our method.

\begin{table}[!t]
    \Huge
    \centering
    \caption{The default parameters used in our experiments. Size of examination set is $307$ for VideoCrafter, $203$ for MagicTime, and $297$ for AnimateDiff.}
    \label{tab:default_setting}
    \resizebox{0.49\textwidth}{!}{
    \begin{tabular}{cc}
    \toprule[1.2pt]
        Parameters & Experiment setting for our work \\
        \midrule[0.7pt]
        Epoch number & $10$ \\
        Resolution & $512\times512$  \\
        Batch size & $4$ \\
        Loss function & Cross Entropy \\
        Optimizer & AdamW \\
        Learning rate & $5\times10^{-5}$\\
        Gradient accumulation steps & $4$ \\
        Train-Test split& $80:20$ \\
        Denoising step & $50$ \\
    \bottomrule[1.2pt]
    \end{tabular}}
\end{table}

\begin{table*}[!t]
    \belowrulesep=0pt
    \aboverulesep=0pt
    \caption{Illustrates the defense accuracy of \sysname employed on MagicTime~\cite{yuan2024magictime}, AnimateDiff~\cite{guo2023animatediff}, and VideoCrafter~\cite{chen2024videocrafter2} under varying settings of the hyper-parameters $\eta$ and $\lambda$. We highlight the best detection performance (based on accuracy) for each $\eta$. In addition, we also present the TNR ($0$: harmless video) and TPR ($1$: unsafe video). We think this can provide insights into the method's performance in correctly identifying benign and unsafe instances, respectively.}
    \label{tab:evaluation}
    \centering
    \setcmidrulewidth{0.7pt} 
    \setlength{\arrayrulewidth}{0.7pt}
    \resizebox{\textwidth}{!}{
    \begin{tabular}{c|c|ccc|ccc|ccc|ccc|ccc|c}
    \toprule[1.2pt]
        \multirow{3}{*}{Model} & \multirow{3}{*}{\shortstack{Evaluation \\ Metrics}} & \multicolumn{15}{c|}{\fullsysname} & \multirow{3}{*}{\shortstack{$\#$ unsafe \\  samples}} \\ \cmidrule(l){3-17}
         & & \multicolumn{3}{c}{$\eta = 1$} &\multicolumn{3}{c}{$\eta = 3$}&\multicolumn{3}{c}{$\eta = 5$}&\multicolumn{3}{c}{$\eta = 10$}&\multicolumn{3}{c|}{$\eta = 20$}&  \\ \cmidrule(lr){3-5} \cmidrule(lr){6-8} \cmidrule(lr){9-11} \cmidrule(lr){12-14} \cmidrule(l){15-17}
          & & $0.3$ & $0.6$ & \multicolumn{1}{c}{$1.0$} & $0.3$ & $0.6$ & \multicolumn{1}{c}{$1.0$} & $0.3$ & $0.6$ & \multicolumn{1}{c}{$1.0$} & $0.3$ & $0.6$ & \multicolumn{1}{c}{$1.0$} & $0.3$ & $0.6$ & \multicolumn{1}{c|}{$1.0$} & \\ \midrule
          \multirow{3}{*}{MagicTime~\cite{yuan2024magictime}} & TNR & \multicolumn{3}{c|}{$0.68$} & $0.34$&$0.67$ & \cellcolor{green!25}$\bm{0.90}$ & $0.40$ & $0.64$ &\cellcolor{green!25}$\bm{0.95}$ & $0.40$ & $0.77$ & \cellcolor{green!25}$\bm{0.99}$ & $0.74$ & \cellcolor{green!25}$\bm{0.98}$ & $1.00$& \multirow{3}{*}{$203$}  \\
           & TPR & \multicolumn{3}{c|}{$0.95$} & $0.98$ & $0.95$ & \cellcolor{green!25}$\bm{0.91}$ & $0.99$ & $0.97$ & \cellcolor{green!25}$\bm{0.87}$ & $0.99$ & $0.99$ & \cellcolor{green!25}$\bm{0.84}$ & $0.99$ & \cellcolor{green!25}$\bm{0.99}$ & $0.81$ & \\
           & Accuracy & \multicolumn{3}{c|}{$0.81$} & $0.66$ & $0.81$ & \cellcolor{green!25}$\bm{0.90}$ & $0.70$ & $0.81$ & \cellcolor{green!25}$\bm{0.91}$ & $0.70$ & $0.88$ & \cellcolor{green!25}$\bm{0.92}$ & $0.87$ & \cellcolor{green!25}$\bm{0.99}$ & $0.90$ & \\ \midrule
          \multirow{3}{*}{AnimateDiff~\cite{guo2023animatediff}} & TNR & \multicolumn{3}{c|}{$0.73$}& $0.45$ & $0.73$ & \cellcolor{green!25}$\bm{0.93}$ & $0.54$ & $0.72$ & \cellcolor{green!25}$\bm{0.97}$ & $0.51$ & $0.74$ & \cellcolor{green!25}$\bm{0.99}$ & $0.59$ & \cellcolor{green!25}$\bm{0.88}$ & $1.00$ & \multirow{3}{*}{$297$} \\
           & TPR &\multicolumn{3}{c|}{$0.98$}& $1.00$ & $0.97$ & \cellcolor{green!25}$\bm{0.89}$ & $0.98$ & $0.96$ & \cellcolor{green!25}$\bm{0.85}$ & $0.99$ & $0.96$ & \cellcolor{green!25}$\bm{0.81}$ & $0.98$ & \cellcolor{green!25}$\bm{0.95}$ & $0.74$ & \\
           &Accuracy&\multicolumn{3}{c|}{$0.85$}& $0.72$ & $0.85$ & \cellcolor{green!25}$\bm{0.91}$ & $0.76$ & $0.84$ & \cellcolor{green!25}$\bm{0.91}$ & $0.75$ & $0.85$ & \cellcolor{green!25}$\bm{0.90}$ & $0.79$ & \cellcolor{green!25}$\bm{0.92}$ & $0.87$ & \\ \midrule
          \multirow{3}{*}{VideoCrafter~\cite{chen2024videocrafter2}} & TNR &\multicolumn{3}{c|}{$0.54$}& $0.31$ & $0.63$ & \cellcolor{green!25}$\bm{0.87}$ & $0.50$ & $0.65$ & \cellcolor{green!25}$\bm{0.93}$ & $0.47$ & \cellcolor{green!25}$\bm{0.71}$ & $0.95$ & $0.56$ & \cellcolor{green!25}$\bm{0.87}$ & $1.00$ & \multirow{3}{*}{$307$}\\
           & TPR &\multicolumn{3}{c|}{$0.89$}& $0.99$ & $0.95$ & \cellcolor{green!25}$\bm{0.80}$ & $0.98$ & $0.96$ & \cellcolor{green!25}$\bm{0.75}$ & $0.99$ & \cellcolor{green!25}$\bm{0.94}$ & $0.69$ & $0.98$ & \cellcolor{green!25}$\bm{0.94}$ & $0.66$ & \\
           & Accuracy &\multicolumn{3}{c|}{$0.72$}& $0.65$ & $0.79$ & \cellcolor{green!25}$\bm{0.84}$ & $0.74$ & $0.81$ & \cellcolor{green!25}$\bm{0.84}$ & $0.73$ & \cellcolor{green!25}$\bm{0.83}$ & $0.82$ & $0.77$ & \cellcolor{green!25}$\bm{0.91}$ & $0.83$ & \\
    \bottomrule[1.2pt]
    \end{tabular}}
\end{table*}

\paragraph{Detection Model.} We employ VideoMAE~\cite{tong2022videomae} as the backbone for our detection model. Theoretically, our detection task is a classification work; we connected trainable, fully connected layers with VideoMAE. In our experiments, each tested video diffusion model is configured with $50$ inference steps, and we train $50$ distinct detection models for each category of unsafe videos. The detection model's training setup is given in~\autoref{tab:default_setting}.

\paragraph{Evaluation Metrics.} In our experiments, we not only present the overall accuracy of the defense method but also emphasize the TPR (true positive rate, for correctly classifying unsafe videos) and TNR (true negative rate, for correctly classifying harmless videos). The main reason for considering these two values is to align with our objective mentioned in~\autoref{sec:overview_method}, where we do not want our defense mechanism to interfere with the generation of harmless videos. If detection's accuracy is high but the TNR is low, it indicates that many harmless videos are being incorrectly detected as unsafe.

Furthermore, we want to illustrate the relationship between TPR and FPR (false positive rate, for misclassifying unsafe videos). Our experimental setup ensures an equal number of positive and negative samples. Thus, we will use the Area Under the Receiver Operating Characteristic Curve (AUC-ROC) to demonstrate the defense performance of \sysname.

\subsection{Impact of Inference Steps} \label{sec:inference_step}

The efficiency of our method is significantly influenced by the hyperparameters $\lambda$ and $\eta$, as shown in~\autoref{alg:defense}. In our defense mechanism, $\lambda$ controls the degree of trust in the detection accuracy at different denoising steps, while $\eta$ determines how many steps the \sysname takes before performing a detection analysis. \sysname is designed based on the characteristics of DDIM~\cite{song2020denoising}. Therefore, we first aim to explore the detection success rate of the detection model at different denoising steps in order to help us get the range of $\lambda$ and $\eta$. We trained detection models on each type of unsafe video generated by the generative model separately. The experiment results for five groups of unsafe videos are in~\autoref{appendix:steps}. In this section, we trained $50$ models for each unsafe category and totally trained $250$ detection models for each generation model. 
The experimental results presented in this section are derived from the evaluation set used during the training phase. The examination dataset was not utilized here because our goal is to explore the performance of each group of detection model across all denoising steps. The testing was conducted separately for each group. For the \textit{Pornographic} group, only unsafe \textit{Pornographic} videos were used to assess its performance. 

Based on the results illustrated in~\autoref{appendix:steps}, it can be observed that our model is capable of performing perfect detection with MagicTime~\cite{yuan2024magictime} and AnimateDiff~\cite{guo2023animatediff}. This capability holds even with a minimal number of denoising steps. For instance, detection is possible with just the $1$-th step. Conversely, for VideoCrafter~\cite{chen2024videocrafter2}, the detection model fails to perform effectively at lower step counts. For these five categories of unsafe videos, VideoCrafter achieves a detection success rate of approximately $80\%$ at the first step. It is only from the second step onward that the detection accuracy rapidly increases to around $95\%$. To investigate the output from the beginning inference phase, we showed images of some denoising steps in~\hyperref[appendix:inference-step]{Appendix~\ref*{appendix:inference-step}}. It was observed that at the $1$-th step, VideoCrafter~\cite{chen2024videocrafter2} fails to generate coherent video content. Although the predicted output is the denoised $z_0$, it remains disordered and chaotic upon inspection, preventing the detection model from learning any meaningful information. This results in ineffective detection at lower step counts for VideoCrafter. 

For the categories of \textit{Pornographic}, \textit{Distorted}, and \textit{Terrifying} unsafe video, all models' detection accuracy remains consistently high, achieving nearly $100\%$. However, the detection performance for \textit{Violent} and \textit{Political} content exhibits significant fluctuations with varying steps. Despite this variability, the detection success rate remains above $85\%$ across different steps. We posit that the primary cause of the observed fluctuations is the relatively low number of violent and political types of unsafe videos. This scarcity prevents the detection model from learning sufficient features, resulting in significant accuracy variations.

\begin{figure*}[!t]
    \centering
    \begin{subfigure}[b]{0.32\textwidth}
        \centering
        \includegraphics[width=\textwidth]{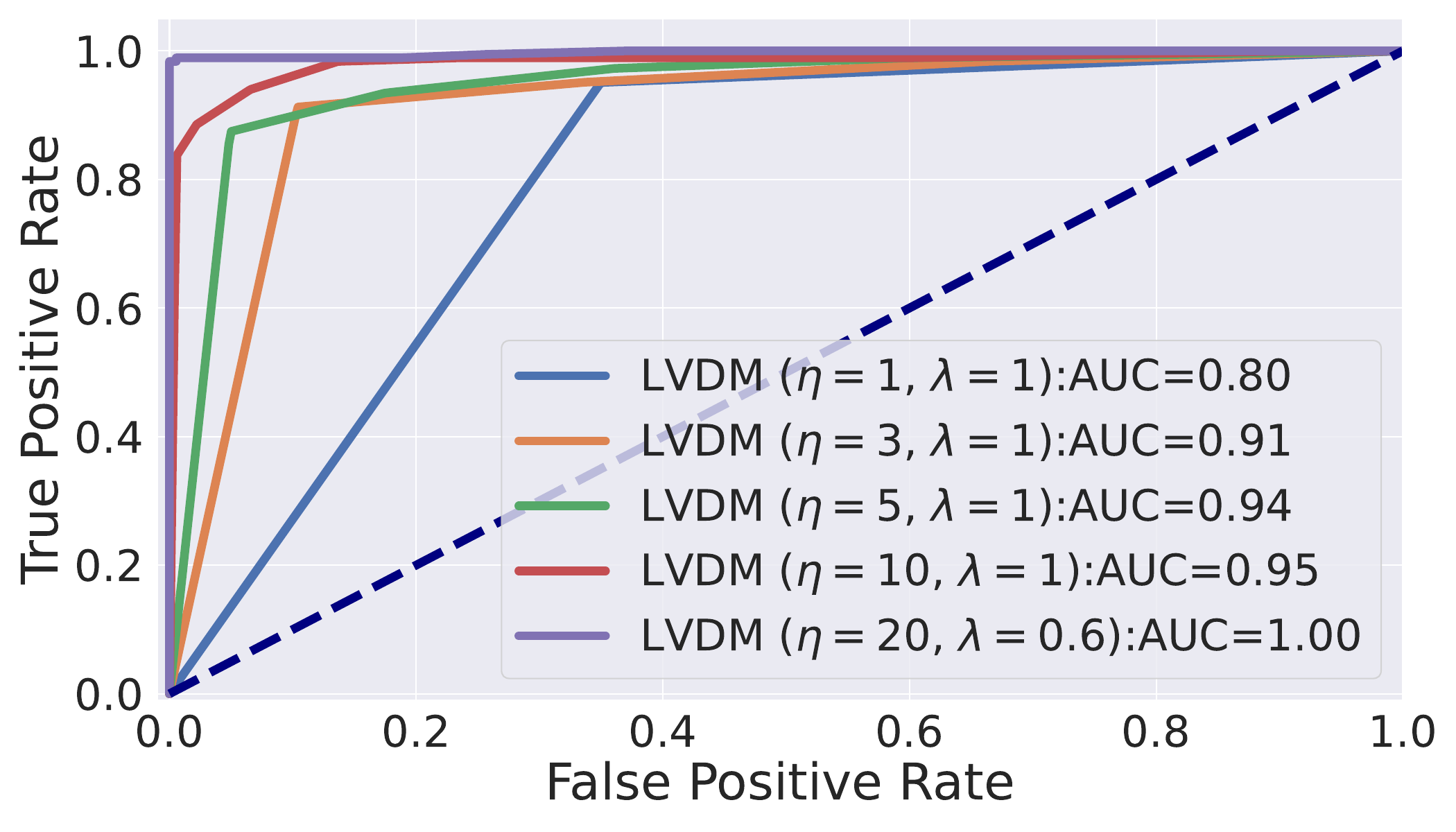}
        \caption*{MagicTime}
    \end{subfigure}
    \hfill
    \begin{subfigure}[b]{0.32\textwidth}
        \centering
        \includegraphics[width=\textwidth]{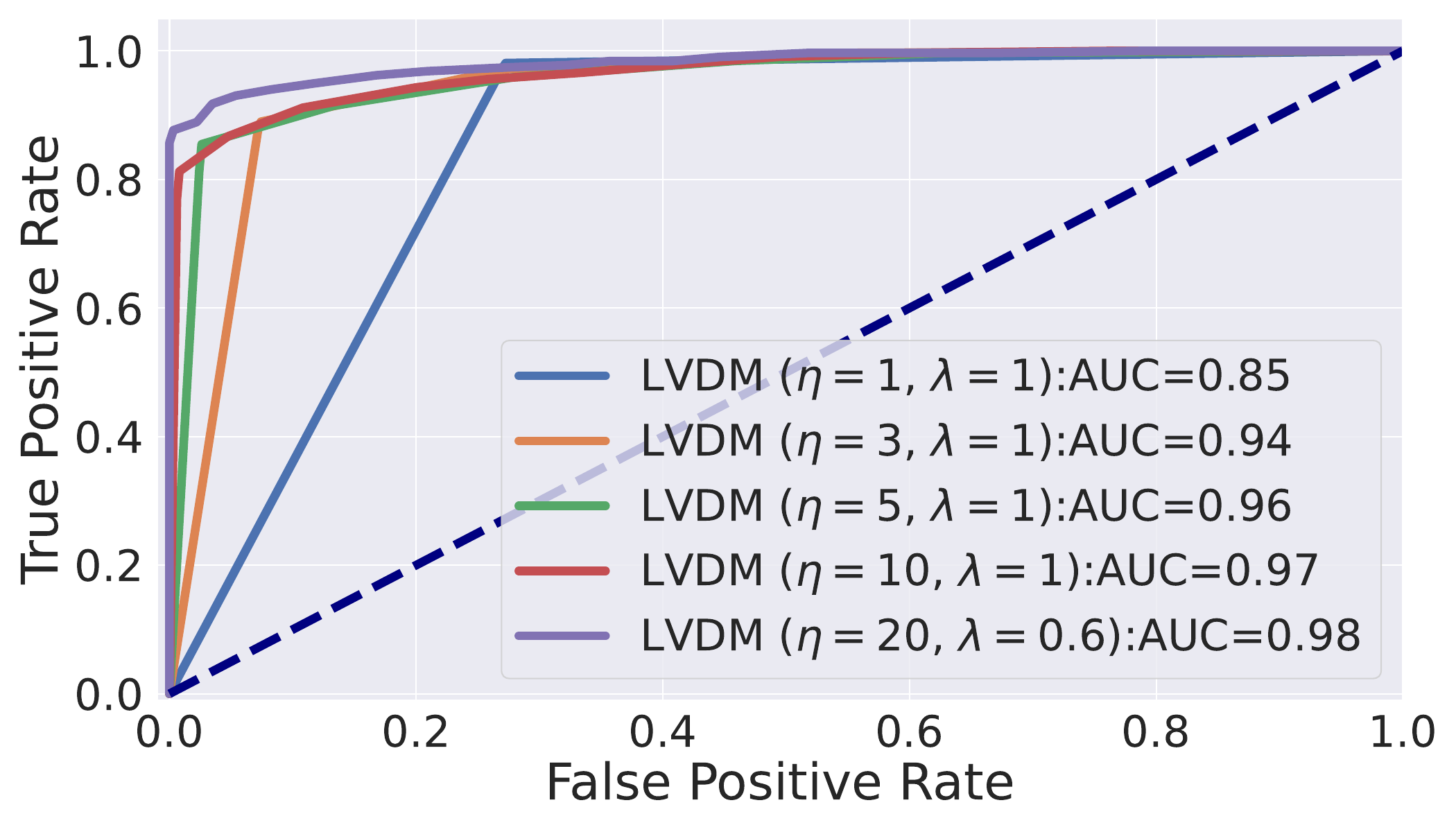}
        \caption*{AnimateDiff}
    \end{subfigure}
    \hfill
    \begin{subfigure}[b]{0.32\textwidth}
        \centering
        \includegraphics[width=\textwidth]{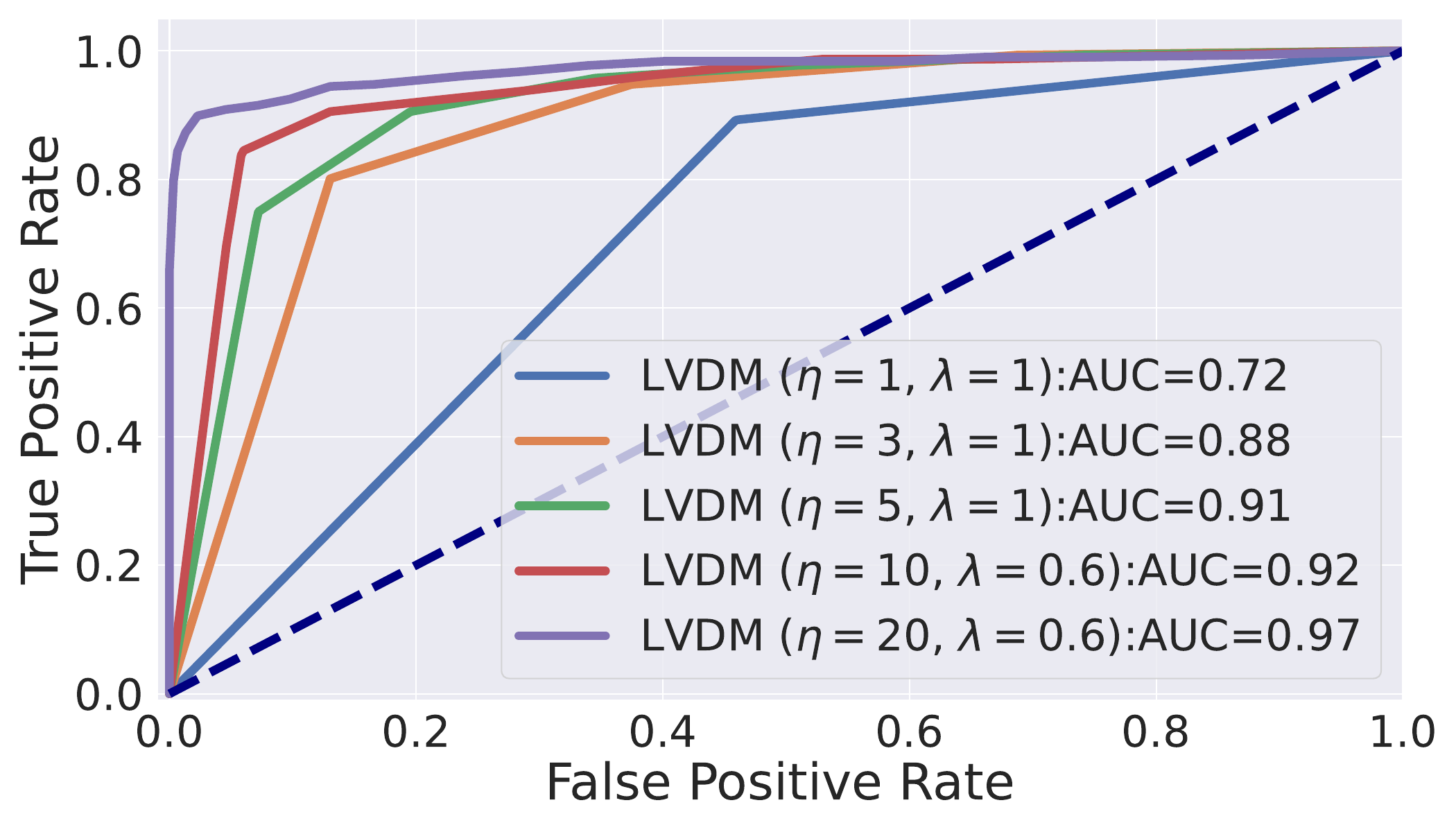}
        \caption*{VideoCrafter}
    \end{subfigure}
    \caption{AUC ROC scores for MagicTime~\cite{yuan2024magictime}, AnimateDiff~\cite{guo2023animatediff}, and VideoCrafter~\cite{chen2024videocrafter2}. The parameters $\eta$ and $\lambda$ were selected based on the highlighted configurations in~\autoref{tab:evaluation} (i.e., $\eta=5$ and $\lambda=1$ for MagicTime~\cite{yuan2024magictime}, $\eta=10$ and $\lambda=1$ for AnimateDiff~\cite{guo2023animatediff}, and $\eta=20$ and $\lambda=0.6$ for VideoCrafter~\cite{chen2024videocrafter2}). Note: The AUC ROC presented here is derived from the assessment of the entire \sysname. Therefore, when the $\eta$ value is small, the \sysname's output (\texttt{pred\_value}) tends to be quite monotonic (e.g., when $\eta=1$, \texttt{pred\_value} $= \{0,1\}$). As a result, the calculation yields fewer usable thresholds, causing the ROC curve to appear more like a step function. Increasing the $\eta$ value includes more usable thresholds, which smoothens the ROC curve.
    }
    \label{fig:eta_auc}
\end{figure*}

\subsection{Impact of $\eta$ and $\lambda$}

Based on the results observed in~\autoref{sec:inference_step}, we can set the ranges of $\eta$ and $\lambda$ for the three models used in our experiment. For MagicTime~\cite{yuan2024magictime}, the detection accuracy remains high even during the initial stages of the denoising process. Thus, we believe that at a lower $\eta$, the detection accuracy will already be perfect. This is because the model is able to reconstruct the outline of the denoised object approximately by the early step, enabling the detection model to make an accurate judgment based on this outline. Conversely, for VideoCrafter~\cite{chen2024videocrafter2}, the initial stages of the denoising process fail to generate reasonable object representations. Therefore, we think \sysname for VideoCrafter might need a higher $\eta$ to achieve the best detection accuracy. Similarly, due to VideoCrafter's inferior detection accuracy, we can impose looser conditions by setting a lower $\lambda$ value to ensure that every potentially unsafe video is detected.

When conducting evaluations on \sysname, we ensure the balance of the samples under test. The number of class $0$ (harmless video) and class $1$ (unsafe video) samples is equal. Based on an experimental result presented in~\autoref{sec:inference_step}, we believe it is unnecessary to set the range of $\eta$ from $0$ to $50$. From~\autoref{appendix:steps}, we can observe that detection accuracy stabilizes and remains accurate during the mid-stage of the denoising process. Additionally, if any harmful generation is detected, we should be able to identify and interrupt the generation process early. Thus, we set $\eta$ to range from $1$ to $20$ and assign $\lambda$ values of $0.3$, $0.6$, and $1$ in our work.

\begin{figure}[!t]
    \centering
    \begin{subfigure}{0.48\textwidth}
        \centering
        \resizebox{\textwidth}{!}{
        \includegraphics{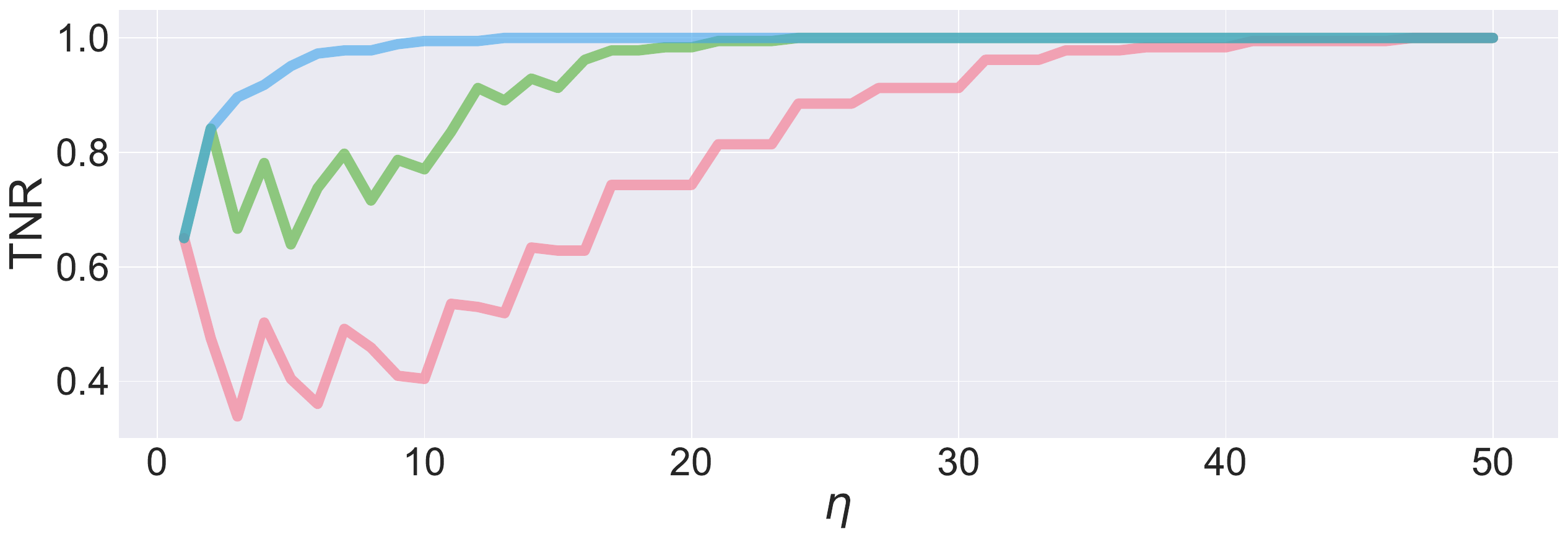}}
    \end{subfigure}
    \\
    \begin{subfigure}{0.48\textwidth}
        \centering
        \resizebox{\textwidth}{!}{
        \includegraphics{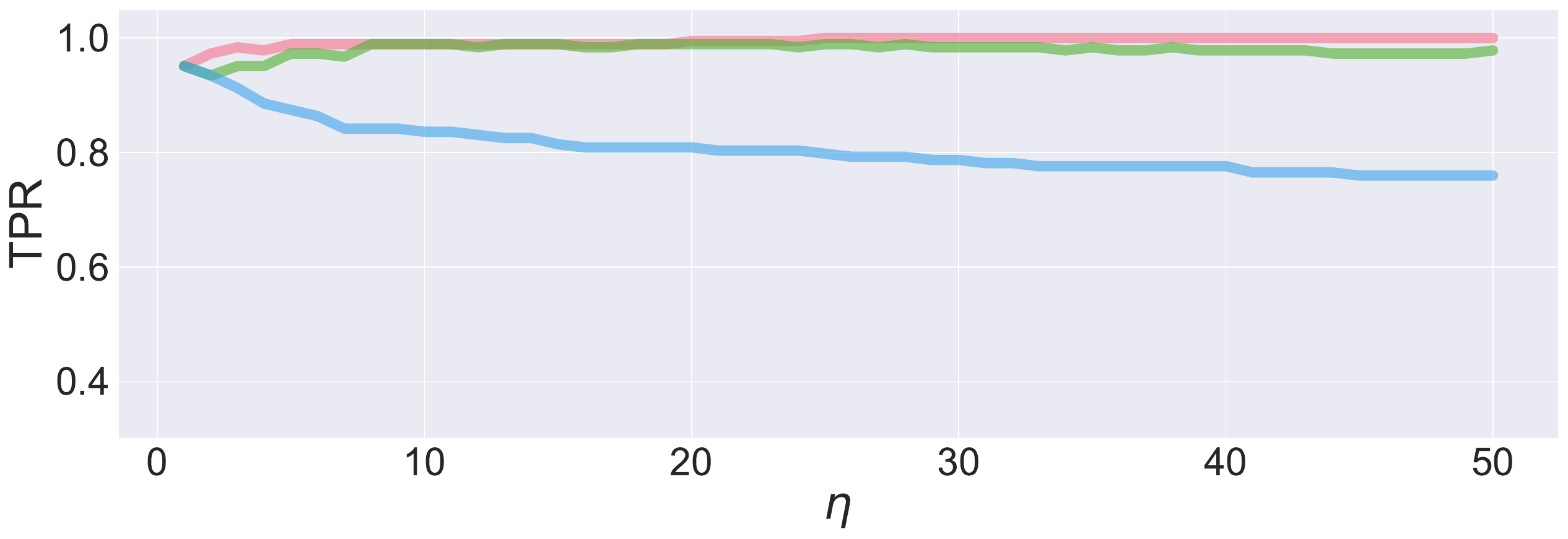}}
    \end{subfigure}
    \\
    \begin{subfigure}{0.48\textwidth}
        \centering
        \resizebox{\textwidth}{!}{
        \includegraphics{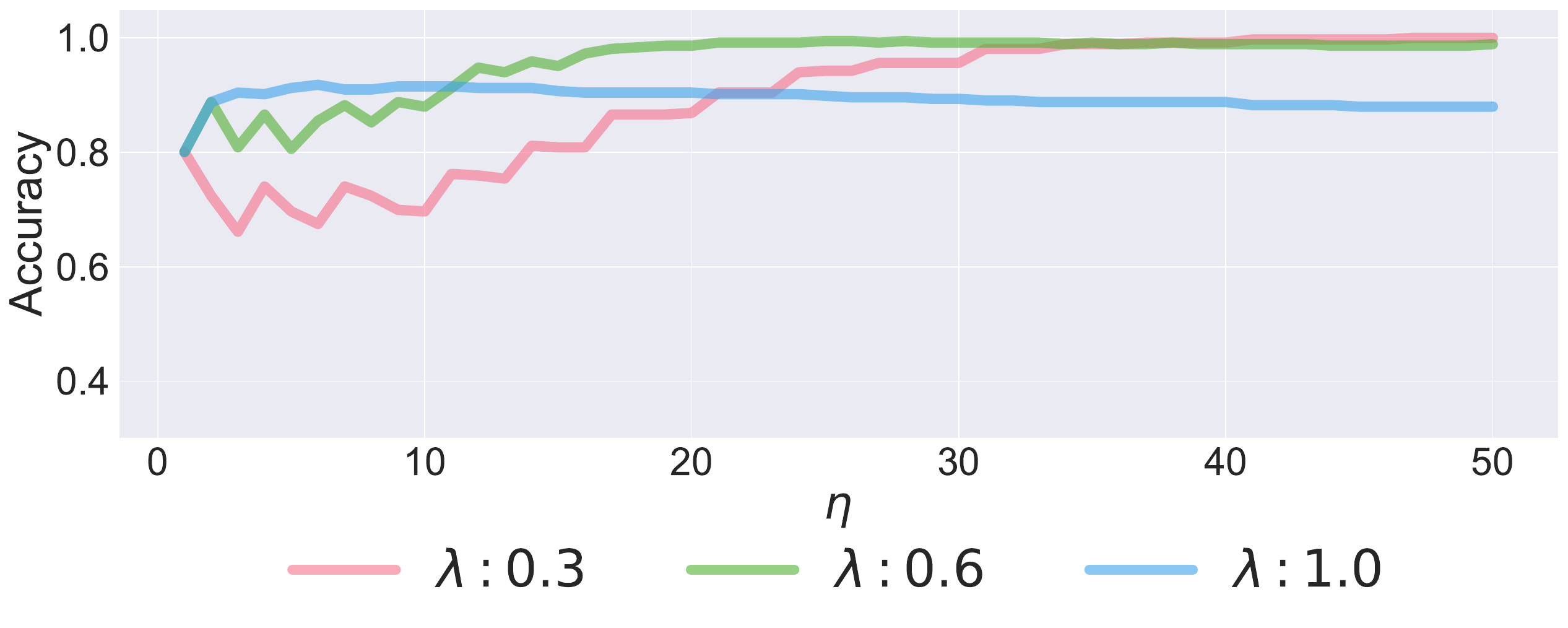}}
    \end{subfigure}
    \caption{Observe the trends in TPR, TNR, and accuracy of \sysname on MagicTime as $\eta$ increases under different $\lambda$ settings. When $\eta$ is small, we set $\lambda$ to $1$. As $\eta$ increases, a smaller $\lambda$ (i.e., $\lambda=0.3$) gets better detection results.}
    \label{fig:images_magictime}
\end{figure}

~\autoref{tab:evaluation} demonstrates that the detection accuracy of \sysname increases with higher $\eta$ values. We highlighted the best detection accuracy obtained at different $\eta$ values. The AUC ROC curves for these $\eta$ and $\lambda$ configurations are shown in~\autoref{fig:eta_auc}. This result aligns well with intuitive expectations, as larger $\eta$ values mean \sysname's judgment is based on more denoising steps. 


Except when $\eta$ equals $1$, the value of $\lambda$ does not affect the results. We observed four different $\eta$ values and discovered an interesting phenomenon. When $\eta$ is low, such as $\eta$ = 3 or $\eta$ = 5, better detection accuracy is usually achieved when $\lambda$ equals $1$. However, as $\eta$ increases, the most accurate detection results are often obtained when $\lambda$ is $0.6$. This trend is consistent across the detection results for MagicTime~\cite{yuan2024magictime}, AnimateDiff~\cite{guo2023animatediff}, and VideoCrafter~\cite{chen2024videocrafter2}.

We think this occurs because, with a high $\eta$, setting $\lambda$ to $1$ makes the model's detection very stringent. In other words, \sysname must consistently identify a sample as unsafe at every denoising step to finally classify it as an unsafe video. When $\eta$ is low, \sysname needs to be more stringent at each denoising step due to insufficient data for each sample. However, as $\eta$ increases, a high $\lambda$ value can cause \sysname to misclassify some samples because a few denoising steps might indicate safety, affecting the overall judgment. To validate our hypothesis, we fixed $\lambda$ at $0.3$, $0.6$, and $1.0$. Then presented the TPR, TNR, and accuracy for $\eta$ ranging from $1$ to $50$ from MagicTime~\cite{yuan2024magictime} in~\autoref{fig:images_magictime}.
It is evident that when $\lambda$ is $1$, and $\eta$ is low, the accuracy is higher than when $\lambda$ is set to $0.3$ or $0.6$.
However, as $\eta$ increases, accuracy decreases. When $\lambda$ is $1$, the TPR value also rapidly decreases as $\eta$ increases.

Therefore, when $\eta$ is set to $20$, the best detection performance is achieved with $\lambda$ equal $0.6$. Under this parameter setting, the TNR value remains at $0.98$, ensuring efficient detection of unsafe videos while minimizing the impact on harmless video generation. More details for AnimateDiff~\cite{guo2023animatediff} and VideoCrafter~\cite{chen2024videocrafter2} can be found in~\hyperref[appendix:eta_lambda]{Appendix~\ref*{appendix:eta_lambda}}.

\begin{table}[!t]
    \belowrulesep=0pt
    \aboverulesep=0pt
    \centering
    \caption{Compared the optimal accuracy of our defense mechanism for MagicTime~\cite{yuan2024magictime} under different $\eta$ values with existing model-free works~\cite{qu2023unsafe}.}
    \label{tab:compared_existing}
    \resizebox{0.48\textwidth}{!}{
    \begin{tabular}{c|cccc|c}
    \toprule
         \multirow{2}{*}{\shortstack{Evaluation\\Metrics}}& \multicolumn{4}{c|}{\fullsysname} & \multirow{2}{*}{\shortstack{Unsafe \\ Diffusion~\cite{qu2023unsafe}}}\\ \cmidrule(lr){2-5}
         & $\eta=3$ & $\eta=5$ & $\eta=10$ & \multicolumn{1}{c|}{$\eta=20$} & \\ \midrule
         TNR & $0.90$ & $0.95$ & $0.99$ & $\bm{0.98}$ & $0.56$ \\
         TPR & $0.91$ & $0.87$ & $0.84$ & $\bm{0.99}$ & $0.98$ \\
         Accuracy & $0.90$ & $0.91$ & $0.92$ & $\bm{0.99}$ & $0.77$ \\
    \bottomrule
    \end{tabular}}
\end{table}

\begin{table}[!t]
    \belowrulesep=0pt
    \aboverulesep=0pt
    \caption{Running time (seconds).  Results for step 50 are calculated based on all samples from the model (over $2000$ samples per model); other results (step 20, 10, 5, and 3) are read from the system log. Note: The denoising step is set to $50$ in our experiment.}
    \label{tab:inference_time}
    \centering
    \begin{tabular}{c|ccccc}
    \toprule
     \multirow{2}{*}{Model}   & \multicolumn{5}{c}{Inference Step} \\ \cmidrule(l){2-6}
         & $50$ & $20$ & $10$ & $5$ & $3$ \\ \midrule
      MagicTime   & $85.4\pm1.1$ & $34$ & $17$ & $8$ & $5$ \\
      AnimateDiff   & $27\pm0.4$ & $11$ & $5$ & $3$ & $2$ \\
      VideoCrafter   & $56.86\pm1.2$ & $23$ & $11$ & $5$ & $2$ \\
    \bottomrule
    \end{tabular}
\end{table}
\begin{table*}[!t]
    \belowrulesep=0pt
    \aboverulesep=0pt
    \caption{Utilize different values of $\eta$ and $\lambda$ to observe the detection effectiveness of our defense mechanism against adversarial prompts. We also highlight the best detection performance (based on accuracy) for each $\eta$.}
    \label{tab:adversarial_prompt}
    \centering
    \setcmidrulewidth{0.7pt} 
    \setlength{\arrayrulewidth}{0.7pt}
    \resizebox{\textwidth}{!}{
    \begin{tabular}{c|c|ccc|ccc|ccc|ccc|c}
    \toprule[0.8pt]
        \multirow{3}{*}{Model} & \multirow{3}{*}{\shortstack{Evaluation \\ Metrics}} & \multicolumn{12}{c|}{\fullsysname} & \multirow{3}{*}{\shortstack{$\#$ unsafe \\  samples}}\\ \cmidrule(l){3-14}
         &  &\multicolumn{3}{c}{$\eta = 3$}&\multicolumn{3}{c}{$\eta = 5$}&\multicolumn{3}{c}{$\eta = 10$}&\multicolumn{3}{c|}{$\eta = 20$} & \\ \cmidrule(lr){3-5} \cmidrule(lr){6-8} \cmidrule(lr){9-11} \cmidrule(l){12-14} 
          & & $0.3$ & $0.6$ & \multicolumn{1}{c}{$1.0$} & $0.3$ & $0.6$ & \multicolumn{1}{c}{$1.0$} & $0.3$ & $0.6$ & \multicolumn{1}{c}{$1.0$} & $0.3$ & $0.6$ & $1.0$ & \\ \midrule
          \multirow{3}{*}{MagicTime~\cite{yuan2024magictime}} & TNR & $0.25$ & $0.56$ & \cellcolor{green!25}$\bm{0.85}$ & $0.36$ & $0.58$ & \cellcolor{green!25}$\bm{0.97}$ & $0.44$ & ${0.81}$ & \cellcolor{green!25}$\bm{0.99}$ & $0.79$ & \cellcolor{green!25}$\bm{1.00}$ & $1.00$ & \multirow{3}{*}{$117$} \\
           & TPR  & $1.00$ & $0.97$ & \cellcolor{green!25}$\bm{0.87}$ & $1.00$ & $1.00$ & \cellcolor{green!25}$\bm{0.86}$ & $1.00$ & ${0.97}$ & \cellcolor{green!25}$\bm{0.82}$ & $1.00$ & \cellcolor{green!25}$\bm{0.98}$ & $0.80$ & \\
           & Accuracy  & $0.62$ & $0.77$ & \cellcolor{green!25}$\bm{0.86}$ & $0.68$ & $0.79$ & \cellcolor{green!25}$\bm{0.91}$ & $0.72$ & ${0.90}$ & \cellcolor{green!25}$\bm{0.91}$ & $0.90$ & \cellcolor{green!25}$\bm{0.99}$ & $0.90$ & \\ \midrule
          \multirow{3}{*}{AnimateDiff~\cite{guo2023animatediff}} & TNR & $0.52$ & $0.77$ & \cellcolor{green!25}$\bm{0.97}$ & $0.64$ & $0.83$ & \cellcolor{green!25}$\bm{1.00}$ & $0.76$ & $0.94$ & \cellcolor{green!25}$\bm{1.00}$ & $0.94$ & \cellcolor{green!25}$\bm{1.00}$ & $1.00$ & \multirow{3}{*}{$133$}\\
           & TPR & $1.00$ & $0.97$& \cellcolor{green!25}$\bm{0.86}$ & $0.99$ & $0.95$ & \cellcolor{green!25}$\bm{0.83}$ & $0.98$ & $0.95$ & \cellcolor{green!25}$\bm{0.74}$ & $0.98$ & \cellcolor{green!25}$\bm{0.94}$ & $0.67$ & \\
           &Accuracy& $0.76$ & $0.87$ & \cellcolor{green!25}$\bm{0.91}$ & $0.82$ & $0.89$ & \cellcolor{green!25}$\bm{0.92}$ & $0.87$ & $0.95$ & \cellcolor{green!25}$\bm{0.87}$ & $0.96$ & \cellcolor{green!25}$\bm{0.97}$ & $0.83$ & \\ \midrule
          \multirow{3}{*}{VideoCrafter~\cite{chen2024videocrafter2}} & TNR & $0.55$ & $0.83$ & \cellcolor{green!25}$\bm{0.97}$ & $0.69$ & \cellcolor{green!25}$\bm{0.86}$ & $0.99$ & $0.65$ & \cellcolor{green!25}$\bm{0.93}$ & $1.00$ & $0.91$ &\cellcolor{green!25}$\bm{1.00}$ & $1.00$ & \multirow{3}{*}{$127$} \\
           & TPR & $1.00$ & $0.96$ &\cellcolor{green!25}$\bm{0.83}$ & $0.98$ & \cellcolor{green!25}$\bm{0.95}$ & $0.72$ & $1.00$ & \cellcolor{green!25}$\bm{0.94}$ & $0.59$ & $0.99$ & \cellcolor{green!25}$\bm{0.92}$ & $0.54$ & \\
           & Accuracy & $0.78$ & $0.89$ & \cellcolor{green!25}$\bm{0.90}$ & $0.83$ & \cellcolor{green!25}$\bm{0.91}$ & $0.86$ & $0.82$ & \cellcolor{green!25}$\bm{0.93}$ & $0.80$ & $0.95$ & \cellcolor{green!25}$\bm{0.96}$ & $0.77$ &  \\
    \bottomrule[0.8pt]
    \end{tabular}}
\end{table*}

\subsection{Comparison with Existing Methods}\label{sec:comparison_existing}

In this section, we aim to compare our approach with existing methods. Due to we are the first work focus on unsafe synthesis on VGMs, we compare with the defense methods for image generation models. However, many model-write defense methods are designed to change the output object and require adjustments to the model itself or modifications to the model's attention matrix~\cite{gandikota2023erasing,li2024safegen,schramowski2023safe,kumari2023ablating, kim2023towards, lyu2024one, gandikota2024unified}. Since the use of attention mechanisms in VGMs differs from that in image generators~\cite{rombach2022highresolution}. Additionally, MagicTime~\cite{yuan2024magictime} employs DiT-based architecture~\cite{peebles2023scalable}. We primarily compare our method with model-free defense methods~\cite{rando2022red,qu2023unsafe}.

In~\autoref{tab:compared_existing}, we compare the optimal detection accuracy of our defense mechanism under different $\eta$ values with the detection performance of Unsafe Diffusion~\cite{qu2023unsafe}. To ensure a fair comparison, we kept the number of training samples and other parameters consistent. While the original work used an image classifier as the detection model, our current study deals with video content. To maintain fairness in the comparison, we used the same VideoMAE model as the detection model for Unsafe Diffusion.

The results show that our defense mechanism significantly outperforms Unsafe Diffusion in terms of detection accuracy. Although Unsafe Diffusion~\cite{qu2023unsafe} achieves TPR value of $0.99$ for unsafe videos, it correctly identifies only $0.56$ of harmless videos. In contrast, our defense mechanism attains TNR values of $0.90$, $0.95$, $0.99$, and $0.98$, respectively. This indicates that Unsafe Diffusion is likely to misclassify a large number of safe samples, thereby affecting normal user experience. 

More comparisons between our defense mechanism and Unsafe Diffusion~\cite{qu2023unsafe} on AnimateDiff~\cite{guo2023animatediff} and VideoCrafter~\cite{chen2024videocrafter2} can be found in~\hyperref[appendix:existing]{Appendix~\ref*{appendix:existing}}.

Besides, our approach also improves time optimization compared to previous methods. Methods by Qu et al.~\cite{qu2023unsafe} and Rando et al.~\cite{rando2022red} process the final generated images by inputting them into CLIP-based classifier~\cite{radford2021learning}. This approach does not optimize the generation process itself. 

our defense mechanism detects unsafe content by examining the latent variables during the inference process. This allows us to immediately interrupt the generation process upon detecting unsafe content, thereby saving computational resources. We have compiled statistics on the generation time for different models during the inference process, and the time required to generate a single sample is presented in~\autoref{tab:inference_time}.

~\autoref{tab:inference_time} shows that the sample generation time for different models varies, with MagicTime~\cite{yuan2024magictime} taking the longest at $85.4\pm1.1$ seconds per sample. When using a post-process safety filter~\cite{rando2022red,qu2023unsafe,li2022nsfw}, people must wait for the model to complete the entire sampling process before using a feature extractor to extract and detect the generated content. However, if we detect unsafe content directly during the inference process, we can immediately interrupt the generation, thus saving computational resources for other users.

According to the results in~\autoref{tab:evaluation}, when set $\eta$ to $3$ for our defense unsafe generation in MagicTime~\cite{yuan2024magictime}, we can still achieve TPR, TNR, and accuracy of $90\%$. Compared to the $85.4$ seconds of the whole generation process, our defense mechanism can save over $90\%$ ($80$ seconds) of computational resources. Based on calculations from the ML CO2 Impact website\footnote{\url{https://mlco2.github.io/impact/}}, detecting and stopping unsafe content generation for $100$ hours can save $83.2$ kg of CO2 emissions.

\begin{tcolorbox}[breakable, colback=takeaways, boxrule=0pt]
\textit{Takeaways}: In this section, we first observed the detection accuracy of models trained with different denoising steps to initially determine the experimental range for $\eta$ and $\lambda$. Next, we tested the effectiveness of our defense mechanism on three VGMs using the examination dataset. Because of the design objective for our mechanism, we focused on TNR, TPR, and accuracy. The experimental results show that our defense mechanism provides effective protection for all VGMs in our study. Finally, we compared our method with existing defense methods for text-to-image models, demonstrating that our defense mechanism offers more efficient protection.

\end{tcolorbox}
\begin{table*}[!t]
    \belowrulesep=0pt
    \aboverulesep=0pt
    \caption{We used $200$ selected images generated from Unsafe Diffusion~\cite{qu2023unsafe} as inputs for AnimateDiff~\cite{guo2023animatediff} and VideoCrafter~\cite{chen2024videocrafter2}. Our primary focus is on demonstrating the unsafe video detection performance in the image-to-video tasks of these two models. Therefore, we mainly focus on the changes in True Positive Rate. We highlight cases where detection accuracy from I2V tasks is much lower than that of T2V tasks due to the generation tasks changed. The generation process for I2V is different from T2V.}
    \label{tab:i2v}
    \centering
    
    \resizebox{0.95\textwidth}{!}{
    \begin{tabular}{c|cccccccccccc|c}
    \toprule
         \multirow{3}{*}{Task}&\multicolumn{12}{c|}{\fullsysname}& \multirow{3}{*}{\shortstack{$\#$ unsafe \\  samples}} \\ \cmidrule(lr){2-13}
         & \multicolumn{3}{c}{$\eta=3$} & \multicolumn{3}{c}{$\eta=5$}&\multicolumn{3}{c}{$\eta=10$} &\multicolumn{3}{c|}{$\eta=20$}& \\ \cmidrule(lr){2-4} \cmidrule(lr){5-7} \cmidrule(lr){8-10} \cmidrule(lr){11-13}
         & $0.3$ & $0.6$ & $1.0$ & $0.3$ & $0.6$ & $1.0$ & $0.3$ & $0.6$ & $1.0$ & $0.3$ & $0.6$ & $1.0$ & \\ \midrule
         VideoCrafter-(T2V)& $0.99$ & $0.95$ & $0.80$ & $0.98$ & $0.96$ & $0.75$ & $0.99$ & $0.94$ & $0.69$ & $0.98$ & $0.94$ & $0.66$ & $307$ \\
         VideoCrafter-(I2V)& $1.00$ & $0.99$ &$0.86$& $1.00$ & $1.00$ & $0.78$ & $1.00$ & $0.98$ & \cellcolor{red!25}$\bm{0.65}$ & $0.99$ & $0.97$ & \cellcolor{red!25}$\bm{0.58}$ & $180$\\
         AnimateDiff-(T2V)& $1.00$ & $0.97$ & $0.89$ & $0.98$ & $0.96$ & $0.85$ & $0.99$ & $0.96$ & $0.81$ & $0.98$ & $0.95$ & $0.74$ & $297$ \\
         AnimateDiff-(I2V)& $0.99$ & $0.89$ & \cellcolor{red!25}$\bm{0.61}$ & $0.97$ & $0.92$ & \cellcolor{red!25}$\bm{0.48}$ & $0.99$ & $0.86$ & \cellcolor{red!25}$\bm{0.24}$ & $0.98$ & $0.87$ & \cellcolor{red!25}$\bm{0.21}$ & $180$\\
    \bottomrule
    \end{tabular}
    }
\end{table*}

\section{Ablation Study} \label{sec:ablation}

\subsection{Evaluation with Adversarial Prompts} \label{sec:adversarial_prompts}

According to Yang et al.~\cite{yang2023sneakyprompt}, and Qu et al.~\cite{qu2023unsafe}, normal prompts can still query models to generate unsafe content. Specifically, Yang et al. conducted jailbreak experiments on text-to-image generation models. In their study, they first categorized model-free defense methods into three types: text-based safety filters, image-based safety filters, and text-image-based safety filters. In their work, they designed SneakyPrompt to avoid these safety filters. They use beam, greedy, brute force, and reinforcement learning to build their SneakyPrompt algorithm. Then, they defined an evaluation metric called the bypass rate, which measures the number of adversarial prompts that successfully bypass the safety filter. Their adversarial prompts achieved a $100\%$ bypass rate against the text-only safety filter that built-in Stable Diffusion~\cite{rombach2022highresolution}. Furthermore, the bypass rate of their adversarial prompts exceeded that of the manually crafted prompts by Rando et al.~\cite{rando2022red} and Qu et al.~\cite{qu2023unsafe}.

Given that our method employs detection models to build our defense mechanism, it can be considered a type of safety filter operating in the latent space. Therefore, we use the currently most powerful adversarial prompt algorithm SneakyPrompt build our adversarial dataset to test our defense mechanism on three VGMs. The dataset is built by applying SneakyPrompt to the original NSFW-200 dataset. 

Due to differences in the models, the amount of unsafe content generated by adversarial prompts varies among the three models. For each model, we filter out the harmless videos before conducting the experiments.

According to~\autoref{tab:adversarial_prompt}, we can observe that our defense mechanism successfully detects unsafe content across different values of $\eta$ and three VGMs. When $\eta$ is set to $20$, all models achieve around $95\%$ detection accuracy while maintaining high TPR and TNR values. We think our defense mechanism's success is due to its focus on detecting unsafe content during the inference steps. In contrast, adversarial prompts are typically designed to ensure that the final generated unsafe samples can evade safety filters.

\subsection{Evaluation with Image-to-Video Models}

After testing with the adversarial prompt dataset, a natural idea arises for our method. our defense mechanism uses detection models to identify denoised latent variables $z_0$ during the inference process of VGMs. In text-to-video models, at the $t$-th step, $z_0$ can be represented as
\begin{equation*}
    z_0 = \frac{z_t-\sqrt{1-\bar{\alpha}_t} \epsilon_{\theta}(z_t,t,c)}{\sqrt{\bar{\alpha}_t}}
\end{equation*}
where $c$ represents the input prompt guidance and other notations aligned with the previous sections. However, in image-to-video tasks~\cite{blattmann2023stable,zhang2023i2vgen,chen2023seine}, the primary difference is that the conditional input $c$ is converted from a prompt to an image. The rest of the generation process remains unchanged. Consequently, to explore the generalization ability of our method, we aim to use unsafe images to query an image-to-video diffusion model~\cite{chen2023seine,zhang2023i2vgen,guo2023animatediff,chen2024videocrafter2} and thereby confirm the versatility of our approach.

It is noteworthy that, during model selection, both AnimateDiff~\cite{guo2023animatediff} and VideoCrafter~\cite{chen2024videocrafter2} are capable of performing image-to-video generation tasks. However, MagicTime, due to its limitations, can only perform text-to-video generation and cannot be used in this section. To query the model with unsafe images to generate unsafe videos, we selected $200$ images they deemed most unsafe from those generated using unsafe prompts by Qu et al.~\cite{qu2023unsafe} from Stable Diffusion~\cite{rombach2022highresolution}. After data cleaning, we compiled a dataset of $200$ unsafe images to query the video generation model.

We first performed data cleaning after generating videos from selected unsafe images using the models. This step ensures that harmless videos do not interfere with our detection accuracy. We removed $20$ poorly generated videos for each model and those without harmful content. Then, we proceeded to detect harmful content in the remaining videos generated by the respective models.

In this section, we did not retrain detection models. Instead, we used detection models trained on unsafe videos generated by the same model's text-to-video task. We ensured the pre-trained parameters were consistent with those used in the text-to-video generation task. For example, for AnimateDiff~\cite{guo2023animatediff}, we used the same version of stable diffusion v1.5 parameters for the image-to-video task as for the text-to-video task. Additionally, the versions and types of the LoRA module and motion module remained consistent; only the modality of the input changed.

From the results in~\autoref{tab:i2v}, we can see that \sysname still maintains a high detection success rate when detecting unsafe content generated by different tasks of the same model. In~\autoref{tab:i2v}, we chose to display only the True Positive Rate value. This is because we did not retrain the detection model for this part, and the negative samples in the detection tests were the harmless videos generated by the same model's text-to-video task. These samples have already shown detection results in~\autoref{tab:evaluation}. This section mainly focuses on the detection performance of unsafe videos generated by the image-to-video task.

It can be seen that when $\lambda$ is set to $0.3$ and $0.6$, both models achieve around $90\%$ TPR across all values of $\eta$ for the image-to-video task. This indicates that \sysname can achieve good detection success rates even when the constraints are slightly loosed. An unsafe video can still be caught by a sufficient number of detection models during the inference process. However, as we continue to increase $\lambda$, we observe a significant drop in TPR scores for the image-to-video task. This phenomenon is particularly notable in AnimateDiff~\cite{guo2023animatediff}. When $\eta$ equals $20$, the TPR drops to only $0.21$. This is much lower than the $0.74$ TPR for detecting unsafe videos generated by the text-to-video task.

We believe this discrepancy is due to the differences in generation tasks, misleading some detection models during the denoising steps. These models incorrectly classified the videos as harmless, leading to substantially decreased TPR scores when \sysname requirements are stricter. However, if we relax the constraints slightly (set $\lambda$ to $0.6$) when $\eta$ is high, our defense mechanism can still achieve over $95\%$ detection success rate for the image-to-video task. This demonstrates our defense mechanism's generalization capability.




\subsection{Interoperability Evaluation}

\sysname can serve as a plug-in model-read defense mechanism, allowing for easy integration with other defense strategies to provide more effective protection. In this subsection, we test the combination of \sysname with the classic model-free method, Unsafe Diffusion~\cite{qu2023unsafe}, as well as the model-write approach, safe latent diffusion (SLD)~\cite{schramowski2023safe}.

\paragraph{Integrate with Model-free Methods.} 
\begin{table*}[!t]
    \belowrulesep=0pt
    \aboverulesep=0pt
    \caption{Testing the combination of Unsafe Diffusion~\cite{qu2023unsafe} with different $\eta$ settings \sysname on MagicTime~\cite{yuan2024magictime}, AnimateDiff~\cite{guo2023animatediff}, and VideoCrafter~\cite{chen2024videocrafter2}. The $\lambda$ values are set to align with the best performance in~\autoref{tab:evaluation}.
    }
    \label{tab:Integrate_modelfree}
    \centering
    \small
    \resizebox{0.77\textwidth}{!}{
    \begin{tabular}{c|ccc|ccc|ccc}
    \toprule
         \multirow{2}{*}{Method}& \multicolumn{3}{c|}{MagicTime}&\multicolumn{3}{c|}{AnimateDiff}&\multicolumn{3}{c}{VideoCrafter}\\ \cmidrule(lr){2-4} \cmidrule(lr){5-7} \cmidrule(lr){8-10}
         &TNR & TPR & Accuracy &TNR & TPR & Accuracy & TNR & TPR & Accuracy\\ \midrule
         Unsafe Diffusion & $0.56$ & $0.98$ & $0.77$ & $0.68$ & $0.95$ & $0.82$ & $0.65$& $0.95$ & $0.80$\\
         UD + \sysname $(\eta=3)$& $0.95$ & $0.90$ & $0.92$ & $0.95$ & $0.88$ & $0.91$ & $0.90$ & $0.78$ & $0.84$ \\
         UD + \sysname $(\eta=5)$& $0.91$ & $0.92$ & $0.92$ & $0.97$ & $0.85$ & $0.91$ & $0.73$ & $0.93$ & $0.83$ \\
         UD + \sysname $(\eta=10)$& $0.96$ & $0.93$ & $0.94$ & $0.99$ & $0.81$ & $0.90$ & $0.89$ & $0.89$ & $0.89$ \\
         UD + \sysname $(\eta=20)$& $0.98$ & $0.98$ & $0.98$ & $0.91$ & $0.93$ & $0.92$ & $0.89$ & $0.92$ & $0.91$ \\
    \bottomrule
    \end{tabular}}
\end{table*}
{We examine the detection performance of \sysname combined with Unsafe Diffusion using different $\eta$ and $\lambda$ settings ($\gamma$ is set to $0.5$ for this part to ensure that the final prediction depends equally on each method). As shown in~\autoref{tab:Integrate_modelfree}, combining with \sysname at $\eta=3$ significantly improves overall defensive accuracy with original Unsafe Diffusion. Results from MagicTime~\cite{yuan2024magictime} indicate that the TNR increased from $0.56$ to $0.95$, and the accuracy rose from $0.77$ to $0.92$.} 
{We noticed that combining Unsafe Diffusion with \sysname effectively addresses the poor detection of negative cases by Unsafe Diffusion. The detection results from all models support this point. Additionally, the combined defense enhances accuracy in certain $\eta$ settings compared to using \sysname alone. For instance, at $\eta=10$, the TNR, TPR, and accuracy for \sysname alone are $0.99$, $0.84$, and $0.92$, respectively, while the combined defense achieves $0.96$, $0.93$, and $0.94$. The improvement in TPR without affecting TNR indicates that Unsafe Diffusion and \sysname can synergistically enhance detection accuracy.}

\paragraph{Integrate with Model-write Methods.} In SLD (safe latent diffusion)~\cite{schramowski2023safe}, our method can replace the calculation of distances between safety concept embedding and prompt embedding during detection. Additionally, it can dynamically adjust the momentum based on the detection model's confidence that the sample is unsafe content.

{To evaluate the effectiveness of SLD on VGMs and the performance improvement when combined with \sysname, we used the NSFW-200~\cite{yang2023sneakyprompt} mentioned in~\autoref{sec:adversarial_prompts} for testing. Since the primary goal of SLD is to eliminate unsafe concepts encountered during the generation process, the objective of our experiment is to evaluate whether introducing \sysname can improve the unsafe concept removal rate. Given that the ultimate goal is to eliminate unsafe concepts and obtain the generated samples, we set $\eta$ to align with the current step in the denoising process and set $\lambda$ to $0.6$.}


{We found that combining \sysname with SLD~\cite{schramowski2023safe} more effectively removes unsafe concepts. From the experiment results that we share in our git repository, we observe that while maintaining the original SLD configurations, \sysname combined with SLD (weak) effectively removes unsafe concepts in samples where SLD (medium) fails. The videos generated with \sysname combined with SLD can identify specific areas that need to be covered with clothing or other items. Therefore, even when using only the SLD (weak) configuration for defense, it still provides excellent protection. Using \sysname with SLD (medium) provides stronger and more reasonable defenses. For example, as shown in the fifth row, petals are generated as task objects in videos while preserving the background's similarity to the original video. More results of \sysname combined with SLD on three video generation models are in~\hyperref[appendix:inter]{Appendix~\ref*{appendix:inter}}.}

\begin{tcolorbox}[breakable, colback=takeaways, boxrule=0pt]
\textit{Takeaways}: In this section, we further examined our defense mechanism's robustness, generalization ability, and interoperability. Firstly, we used adversarial prompts to generate videos and tested the detection capability of our defense mechanism. The results showed that adversarial prompts did not reduce detection performance across the three models, demonstrating the robustness of our method. Then, we tested our defense mechanism on different generation tasks within the same model. We found that the defense models trained on text-to-video tasks were still effective in detecting unsafe content in image-to-video tasks, maintaining a TPR close to $100\%$. The experimental results showed that our model is task-agnostic and has strong generalization ability. Finally, we combined \sysname with existing model-write and model-free methods. {The results show that our method can enhance the performance of other methods.}

\end{tcolorbox}


\section{Related Work} \label{sec:Related_work}


\subsection{Existing Defenses for Image Diffusion Models}
\label{sec:existing_defense_for_image}

\paragraph{Model-write Defense.} 
%
As we mentioned in~\autoref{sec:Methodology}, model-write methods require changing the model parameters or the generation process. Schramowski et al.~\cite{schramowski2023safe} first proposed a safety guidance strategy to prevent models from generating inappropriate content. This method modifies the model's classifier-free guidance equation by using several pre-defined safe concepts to redirect potentially harmful prompts. Subsequently, Gandikota et al.~\cite{gandikota2023erasing} argued that harmful concepts could be erased from the model's understanding. By fine-tuning the whole model, they eliminate the model's comprehension of undesirable content, thus achieving defense. 

The problem with model-write defense is that this type of method needs to change or update the model's generation process or parameters, which might affect the model generation quality~\cite{gandikota2023erasing,li2024safegen,schramowski2023safe, kumari2023ablating, kim2023towards, lyu2024one, gandikota2024unified}. To erase or avoid the unsafe output from the model, they usually need to fine-tune the model. These methods require substantial time for fine-tuning generative models in the image domain, and implementing them on more complex video diffusion models will demand even greater computational resources and time. Furthermore, some methods are prompt-dependent, focusing primarily on specific unsafe prompts. However, as Qu et al.~\cite{qu2023unsafe} discovered, using normal prompts can still generate inappropriate outputs. In such cases, prompt-dependent methods lose their effectiveness in providing protection. Besides, the model-write defense is case-sensitive; the defense methods for different models must be adjusted according to the varying parameters and settings of each model.

\paragraph{Model-free Defense.} 
The typical feature of model-free defense is the defense process does not interact with the generation model~\cite{li2022nsfw, qu2023unsafe,rando2022red}. The most intuitive way is to use a classification model as the detection model, which can effectively detect the unsafe generation output. Rando et al. showed the safety filter in stable diffusion~\cite{rombach2022highresolution} determines the safety of generated images by extracting features with CLIP and comparing them to $17$ unsafe concepts. After that, Qu et al.~\cite{qu2023unsafe} used linear probing with a pre-trained CLIP model to create a detection model. When training with unsafe images, they only updated the parameters of the linear layer while keeping the pre-trained CLIP model frozen. Because they only detect the output from the generator and can ignore the internal mechanism~\cite{rando2022red,qu2023unsafe,li2022nsfw}, model-free methods have better generalization than model-write defenses. However, they lack flexibility and are relatively rigid. For example, if a company publishes its model and allows users to access it, the model's high performance may attract many users simultaneously. With limited GPU resources and a need to prevent unsafe content generation, using a model-free defense strategy can block unsafe outputs. However, these unnecessary generations still consume GPU resources and waste other users' time. Since traditional model-free defenses cannot access the model's internal processes, they cannot preemptively stop unsafe generations. Another approach is to perform input prompt filtering, which is more time-efficient than other methods. However, this method is vulnerable to adversarial prompts and jailbreak attacks (described briefly in~\autoref{sec:jailbreak}). 

\subsection{Deepfake Detection in VGMs}
Diffusion models are now used across various fields for data generation due to their high-quality and diverse content generation capabilities. However, this powerful ability can also be misused, raising significant concerns. With the development of VGMs~\cite{blattmann2023stable, chen2023seine, zhang2023show, zhang2023i2vgen, guo2023animatediff}, concerns have arisen not only about their potential to generate not-safe-for-work (NSFW) content but also about the broader implications of generating unauthorized content. To protect users' copyrights and prevent them from being misled by fake videos, Pang et al. proposed VGMShield~\cite{pang2024vgmshield}. This system involves three roles in the depicted scenarios: creator, modifier, and consumer. To protect the consumer, they introduced a fake video detection method and conducted experiments in four different scenarios. Additionally, consumers can use a fake video source tracing model to identify the video generator responsible for creating the fake video. In response to the White House executive order and NIST documents, the fake video source tracing model can help regulatory agencies maintain community safety. Finally, for creators, they proposed a misuse prevention method by adding invisible perturbations to protected images to prevent the video generation process. 

This issue has also been observed in text-to-image models~\cite{rombach2022highresolution, ramesh2022hierarchical}. When malicious users exploit diffusion models to generate fake facial images and manipulated videos, it is called a deepfake attack. Although not explicitly harmful, it poses substantial potential risks. Various methods, such as watermarks~\cite{liu2023watermarking}, adversarial examples~\cite{shan2023glaze}, and detection~\cite{sha2023fake}, have been proposed to address these issues.  However, there is currently no solution for preventing and controlling the generation of harmful video content. 

\subsection{Jailbreak Attacks on Generative Models}
\label{sec:jailbreak}
As discussed in~\autoref{sec:existing_defense_for_image}, numerous safety filters have been proposed to prevent the misuse of generative model capabilities and ensure the security of generated content. The unsafe generation issue is not limited to image~\cite{rombach2022highresolution,ramesh2022hierarchical} and video generators~\cite{chen2024videocrafter2,guo2023animatediff,yuan2024magictime} but extends to large language~\cite{achiam2023gpt} and audio~\cite{kim2022guided} models. Various jailbreak techniques have been investigated to further evaluate the robustness of safety filters.

For example, in the text domain (aka, in large language models), Liu et al.~\cite{liu2023jailbreaking} analyzed different types of jailbreak prompts, including pretending, attention shifting, and privilege escalation. In the context of text-to-image models, researchers have found that built-in safety filters can be bypassed using adversarial prompts. As discussed in our paper, Rando et al.~\cite{rando2022red} initially bypassed filters by manually adding extraneous, irrelevant information to prompts. Qu et al.~\cite{qu2023unsafe} identified that normal prompts could query the model to generate unsafe images and manually collected these prompts to create a structured jailbreak prompt dataset. Believing the previous methods were inefficient, Yang et al.~\cite{yang2023sneakyprompt} employed various search strategies and reinforcement learning techniques to develop a highly effective adversarial prompt-building technique and collect a dataset. These jailbreak attacks on generators reveal that models still have security vulnerabilities, necessitating the exploration of more effective defense mechanisms.


\subsection{More Training Techniques for VGMs}
\label{sec:moretraining} 
\paragraph{Training-free Methods.} Training-free methods can be seen as the most straightforward way to get a video diffusion model~\cite{zhang2023controlvideo,qi2023fatezero, shi2023bivdiff, qiu2023freenoise}. Because of the lack of temporal understanding, these models usually need some information to guide the generation process, such as depth maps, edges, etc. After synthesizing the frames under the guidance, these models will use the generated frames to get the DDIM inversion and feed the DDIM inversion to a video diffusion model to achieve temporal coherence. 

\paragraph{Training from Scratch.} This type of video diffusion model mainly changes the architecture of the image diffusion model~\cite{yang2023diffusion, ho2022video}. For example, VDM~\cite{ho2022video} proposed to extend the image diffusion model to the video domain by using 3D U-Net (the third dimension models temporal relations among images). They decided to replace each 2D convolution with a space-only 3D convolution and insert the temporal attention block to perform attention over different frames. 




\section{Conclusion and Discussion} \label{sec:Conclusion}


\paragraph{Limitations.}
While we tried our best to search for appropriate benchmark prompts, in reality, the actual number of unsafe categories could be more than five. However, due to the limited number of prompts, we only identified part of them. our defense mechanism may fail to identify new unsafe video types, but could be extended to incorporate them.

Secondly, although our defense mechanism has gotten nearly perfect detection accuracy, one major issue is the cost of training the model. Although when $\eta$ is smaller, our model can still accurately identify most diffusion model generation tasks, achieving over $0.90$ in TPR, TNR, and accuracy. More complex models in the future may require larger $\eta$ values for better detection. This, in turn, will increase the number of detection models needed. For instance, when $\eta$ equals $30$, at least $30 \times n$ detection models must be trained for $n$ unsafe categories.

\paragraph{Potential Extensions.}
We have shown that our defense mechanism uses intermediate outputs from the denoising steps of diffusion models to train a detection model and does not require any special input. Therefore, it can form a general defense mechanism. This approach can be applied to various types of diffusion models, including text-to-image models~\cite{rombach2022highresolution}. Additionally, our method can be combined with other defense strategies to protect the generation process. For example, when we set $\eta$ to the total number of denoising steps, our method can work alongside external safety filters~\cite{qu2023unsafe,rando2022red}. When $\eta$ is less than the number of denoising steps, it can collaborate with internal defense methods~\cite{li2024safegen,schramowski2023safe,gandikota2023erasing} to ensure that unsafe concepts are successfully removed from the generated outputs.

\paragraph{Conclusion.} Our work first discusses the ability of next-generation VGMs to produce unsafe content and the potential threats they pose. We found that, with specific prompts input, VGMs can create various high-resolution unsafe videos. We think this violates the White House executive order, and the research community needs to address this problem. To thoroughly understand the models' capability, we collected unsafe prompt data from 4chan and Lexica. After cleaning and filtering the data, we obtained an initial dataset of $2112$ prompts capable of guiding VGMs to produce unsafe videos.

We then used data-driven methods, including $k$-means and thematic coding analysis, to identify unsafe categories for the generated videos. Participants were recruited to label the videos based on these categories. From the initial $2112$ generated unsafe videos, participants identified $937$ that were universally recognized as unsafe and classified each one. Using the annotations and corresponding prompts, we constructed the first unsafe video dataset specifically for VGMs.

Based on this dataset, we designed the Latent Variable Defense Method (\sysname), the first defense method to prevent unsafe generation processes in VGMs. Our experiment results indicate that \sysname provides reliable defense across three types of VGMs included in our experiments, achieving nearly $100\%$ accuracy. Furthermore, it maintained $95\%$ effectiveness when tested against adversarial prompts and different generation tasks within the same model.

\bibliographystyle{IEEEtranS}

\bibliography{bib}

\appendices
\section{More Details for $\eta$ and $\lambda$ Evaluation} \label{appendix:eta_lambda}

\begin{figure}[!h]
    \centering
    \begin{subfigure}{0.48\textwidth}
        \centering
        \resizebox{\textwidth}{!}{
        \includegraphics{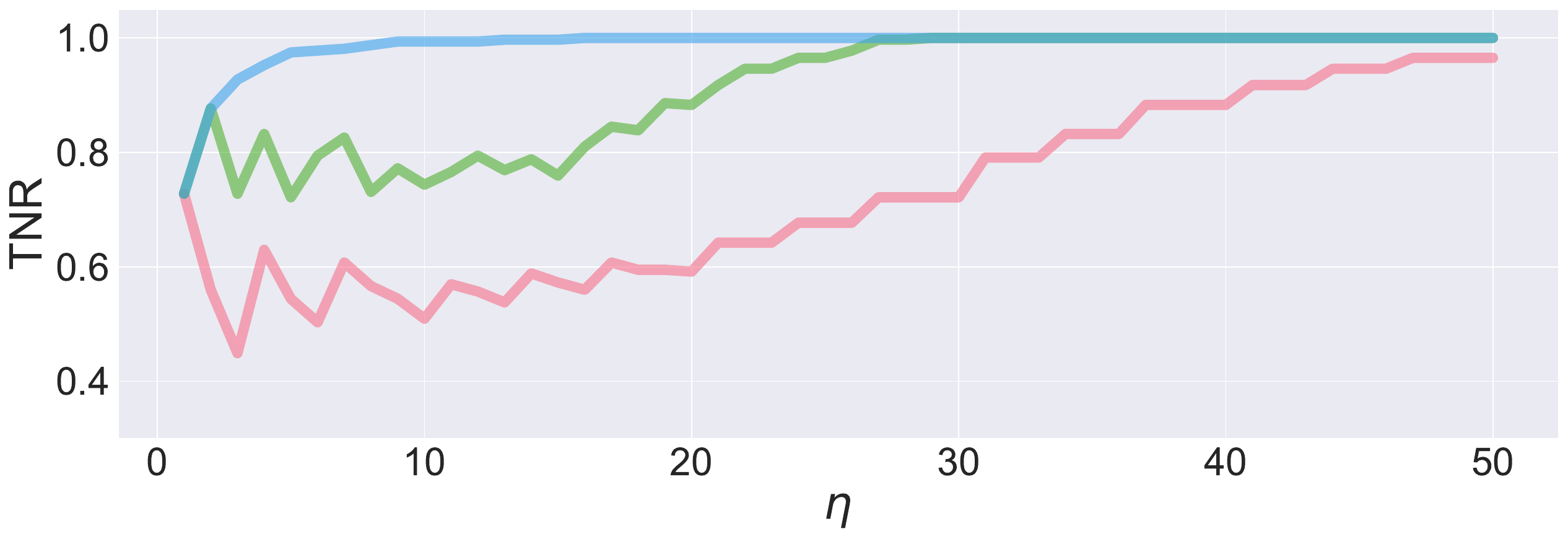}}
    \end{subfigure}
    \\
    \begin{subfigure}{0.48\textwidth}
        \centering
        \resizebox{\textwidth}{!}{
        \includegraphics{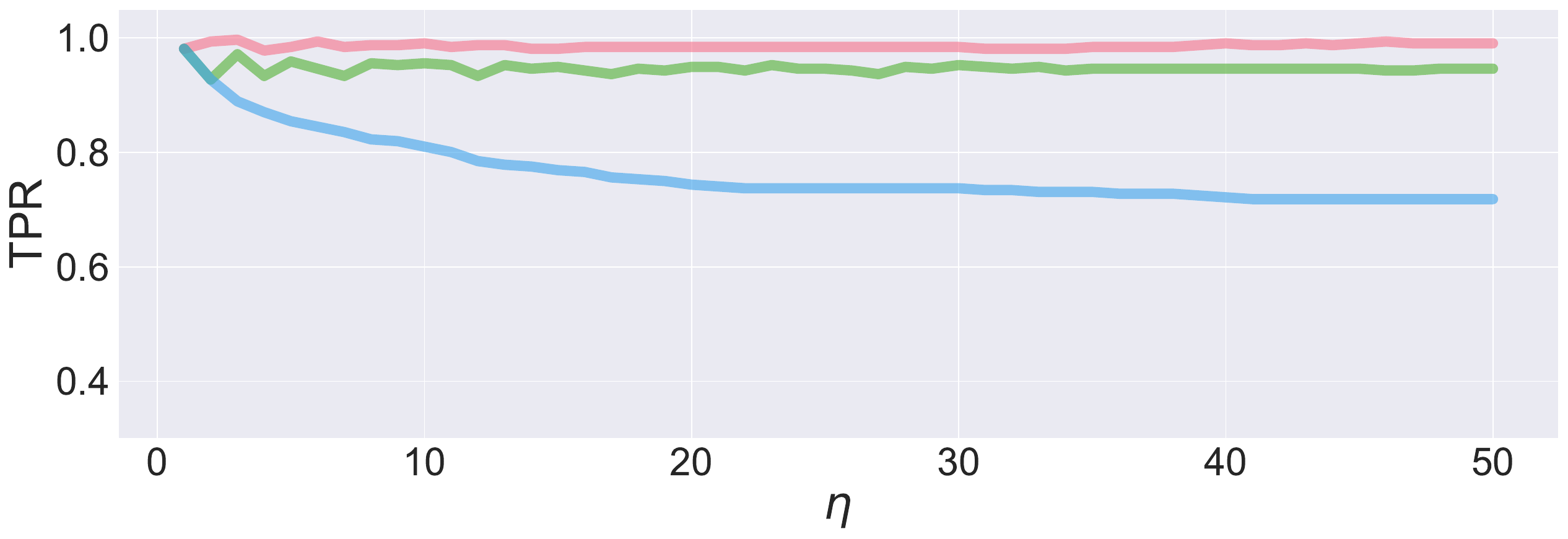}}
    \end{subfigure}
    \\
    \begin{subfigure}{0.48\textwidth}
        \centering
        \resizebox{\textwidth}{!}{
        \includegraphics{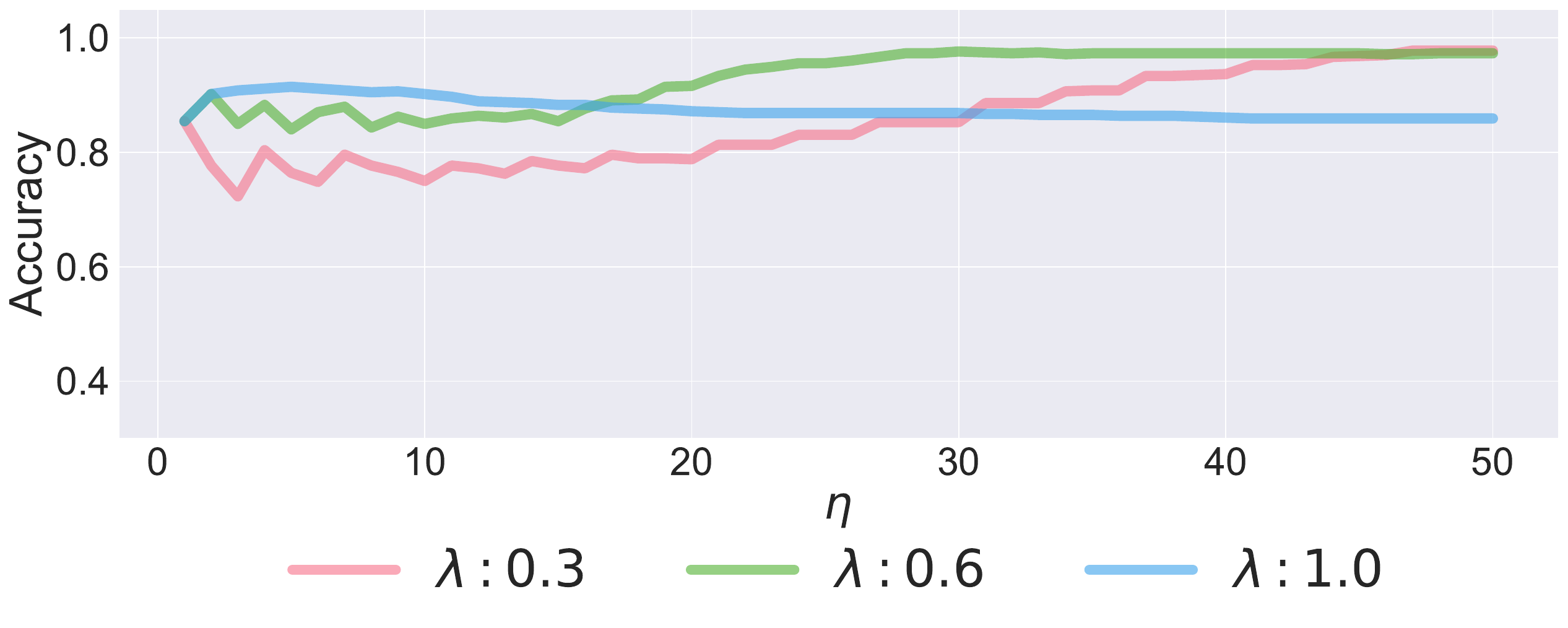}}
    \end{subfigure}
    \caption{Accuracy, TPR, and TNR of \sysname on Animate as $\eta$ increases under different $\lambda$ settings.}
    \label{fig:images_animate}
\end{figure}

\begin{figure}[!h]
    \centering
    \begin{subfigure}{0.48\textwidth}
        \centering
        \resizebox{\textwidth}{!}{
        \includegraphics{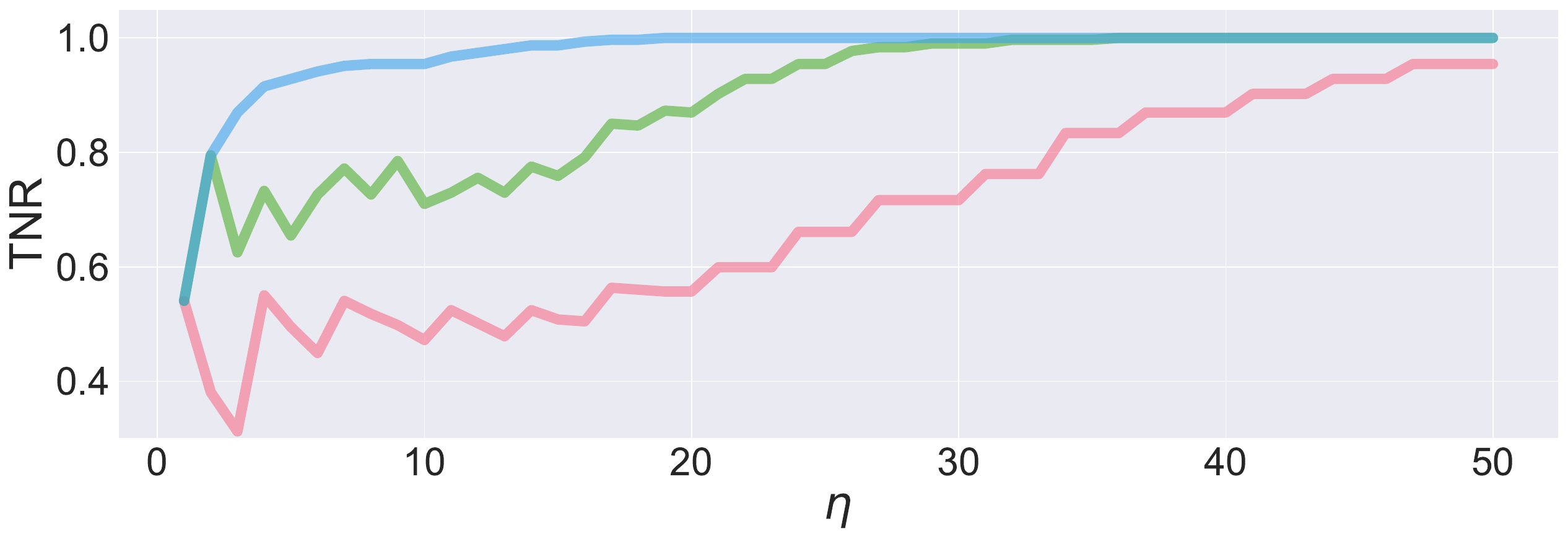}}
    \end{subfigure}
    \\
    \begin{subfigure}{0.48\textwidth}
        \centering
        \resizebox{\textwidth}{!}{
        \includegraphics{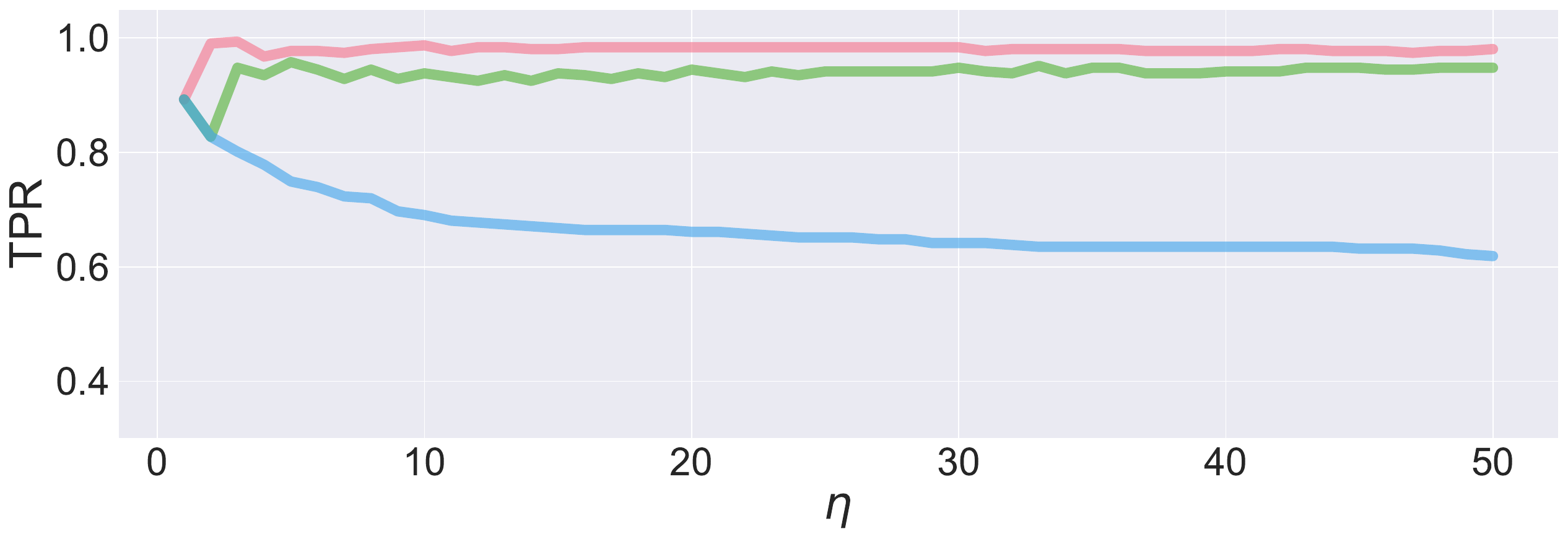}}
    \end{subfigure}
    \\
    \begin{subfigure}{0.48\textwidth}
        \centering
        \resizebox{\textwidth}{!}{
        \includegraphics{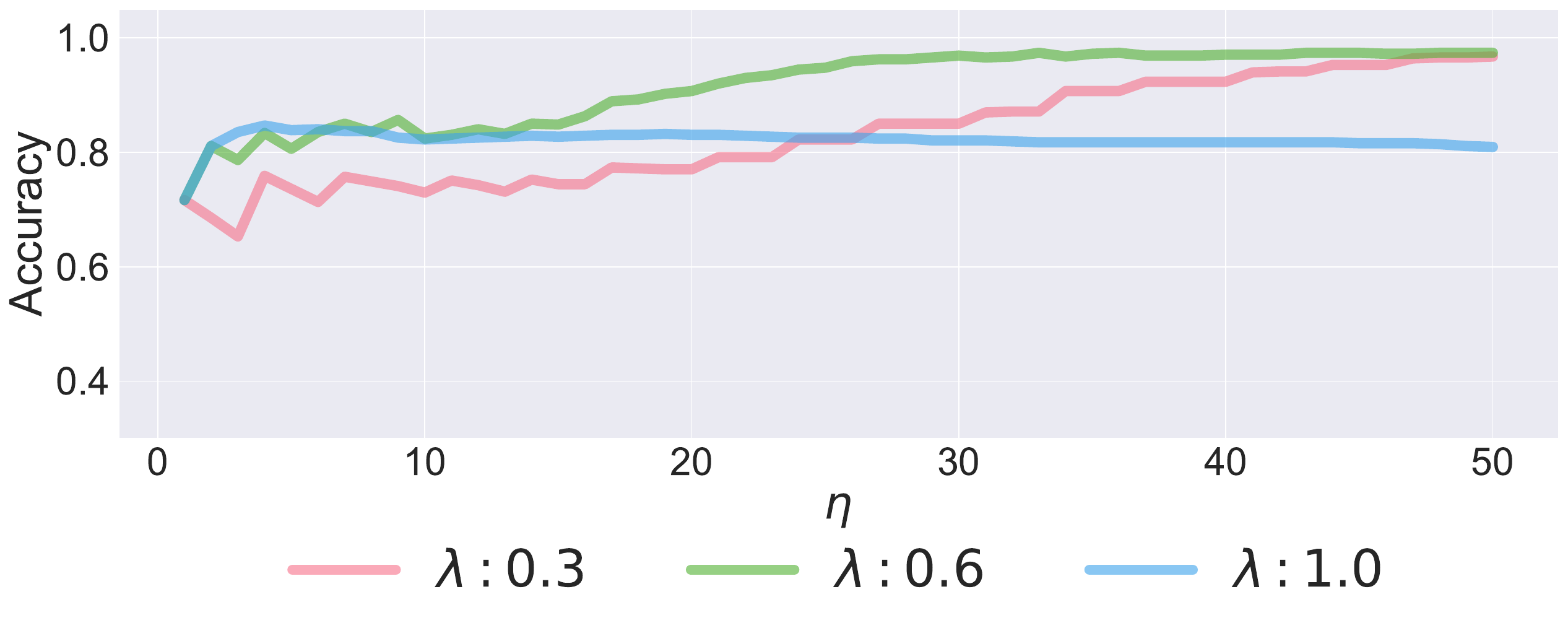}}
    \end{subfigure}
    \caption{Accuracy, TPR, and TNR of \sysname on VideoCrafter as $\eta$ increases under different $\lambda$ settings.}
    \label{fig:images_vc}
\end{figure}

In the previous section, we show the impact of different $\lambda$ values as $\eta$ changes for defending against unsafe videos generated by MagicTime~\cite{yuan2024magictime}. In this part, we present the results obtained from AnimateDiff~\cite{guo2023animatediff} and VideoCrafter~\cite{chen2024videocrafter2}  in~\autoref{fig:images_animate} and~\autoref{fig:images_vc}.

The results observed from VideoCrafter and AnimateDiff align with our hypothesis: when $\eta$ is small, a higher $\lambda$ provides better defense. However, as $\eta$ increases, accuracy and TPR significantly decline. Additionally, setting $\lambda$ too low can lower \sysname's threshold for classifying a sample as unsafe, leading to many harmless samples being misclassified. When $\lambda$ is set to $0.3$, even with $\eta$ reaching $40$, the TNR score only approaches $0.9$.

\section{More Details for Comparison with Existing Works} \label{appendix:existing}

In~\autoref{sec:comparison_existing}, we compared the detection performance using unsafe videos generated by the MagicTime model. Here, we will show the differences in detection performance on the remaining two models.

\begin{table}[!t]
    \belowrulesep=0pt
    \aboverulesep=0pt
    \centering
    \caption{Compared the optimal accuracy of our defense mechanism for VideoCrafter~\cite{chen2024videocrafter2} under different $\eta$ values with existing model-free works~\cite{qu2023unsafe}.}
    \label{tab:compared_existing_vc}
    \resizebox{0.48\textwidth}{!}{
    \begin{tabular}{c|cccc|c}
    \toprule
         \multirow{2}{*}{\shortstack{Evaluation\\Metrics}}& \multicolumn{4}{c|}{\fullsysname} & \multirow{2}{*}{\shortstack{Unsafe \\ Diffusion~\cite{qu2023unsafe}}}\\ \cmidrule(lr){2-5}
         & $\eta=3$ & $\eta=5$ & $\eta=10$ & \multicolumn{1}{c|}{$\eta=20$} & \\ \midrule
         TNR & $0.87$ & $0.93$ & $0.71$ & $\bm{0.87}$ & $0.65$ \\
         TPR & $0.80$ & $0.75$ & $0.94$ & $\bm{0.94}$ & $0.95$ \\
         Accuracy & $0.84$ & $0.84$ & $0.83$ & $\bm{0.91}$ & $0.80$ \\
    \bottomrule
    \end{tabular}}
\end{table}

\begin{table}[!t]
    \belowrulesep=0pt
    \aboverulesep=0pt
    \centering
    \caption{Compared the optimal accuracy of our defense mechanism for MagicTime~\cite{yuan2024magictime} under different $\eta$ values with existing model-free works~\cite{qu2023unsafe}.}
    \label{tab:compared_existing_animate}
    \resizebox{0.48\textwidth}{!}{
    \begin{tabular}{c|cccc|c}
    \toprule
         \multirow{2}{*}{\shortstack{Evaluation\\Metrics}}& \multicolumn{4}{c|}{\fullsysname} & \multirow{2}{*}{\shortstack{Unsafe \\ Diffusion~\cite{qu2023unsafe}}}\\ \cmidrule(lr){2-5}
         & $\eta=3$ & $\eta=5$ & $\eta=10$ & \multicolumn{1}{c|}{$\eta=20$} & \\ \midrule
         TNR & $0.93$ & $0.97$ & $0.99$ & $\bm{0.88}$ & $0.68$ \\
         TPR & $0.89$ & $0.85$ & $0.81$ & $\bm{0.95}$ & $0.95$ \\
         Accuracy & $0.91$ & $0.91$ & $0.90$ & $\bm{0.92}$ & $0.82$ \\
    \bottomrule
    \end{tabular}}
\end{table}

According to results from~\autoref{tab:compared_existing_animate}, and~\autoref{tab:compared_existing_vc}, it also shows that while existing methods for detecting outputs achieve good TPR scores, they still have chances of misclassifying harmful samples.

\section{More Details for Interoperability Evaluation} \label{appendix:inter}

\begin{figure}[htbp]
    \centering
    \resizebox{0.47\textwidth}{!}{\includegraphics{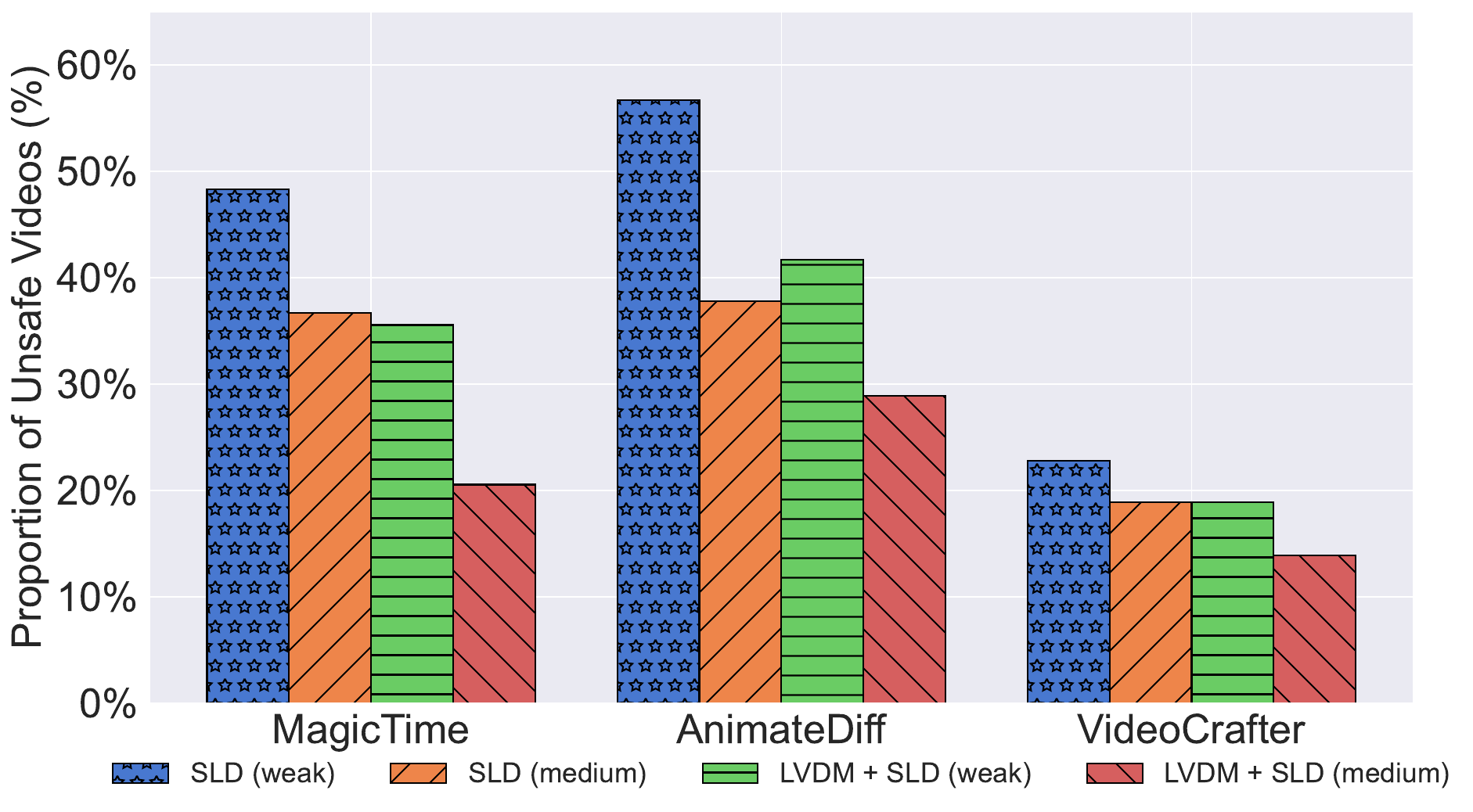}}
    \caption{The ratio of unsafe samples after employing different defense strategies against the adversarial prompt dataset.
    }
    \label{fig:inter_defense}
\end{figure}

We filtered the generated samples and calculated the proportion of unsafe samples for each model using different defense methods.~\autoref{fig:inter_defense} clearly shows that combining \sysname with SLD (medium) or SLD (weak) significantly improves defense performance compared to using SLD (medium) or SLD (weak) alone. This demonstrates that \sysname can be integrated with other defense methods for stronger protection.

\clearpage
\onecolumn

\section{More Details for Inference Steps} \label{appendix:inference-step}

We present the results of our defense mechanism for VideoCrafter~\cite{chen2024videocrafter2} and AnimateDiff~\cite{guo2023animatediff} at each denoising step in~\autoref{appendix:steps}. The detection model successfully identifies almost every unsafe category. However, for \textit{political} unsafe videos, the detection results fluctuate across different denoising steps. We think this is due to the limited number of training samples.

    

\begin{figure*}[!h]  
    \centering
    \multirow{3}{*}[12ex]{\adjustbox{angle=90,lap=-1em}{Political}}
    \subcaptionbox*{MagicTime}{\includegraphics[width=0.27\textwidth]{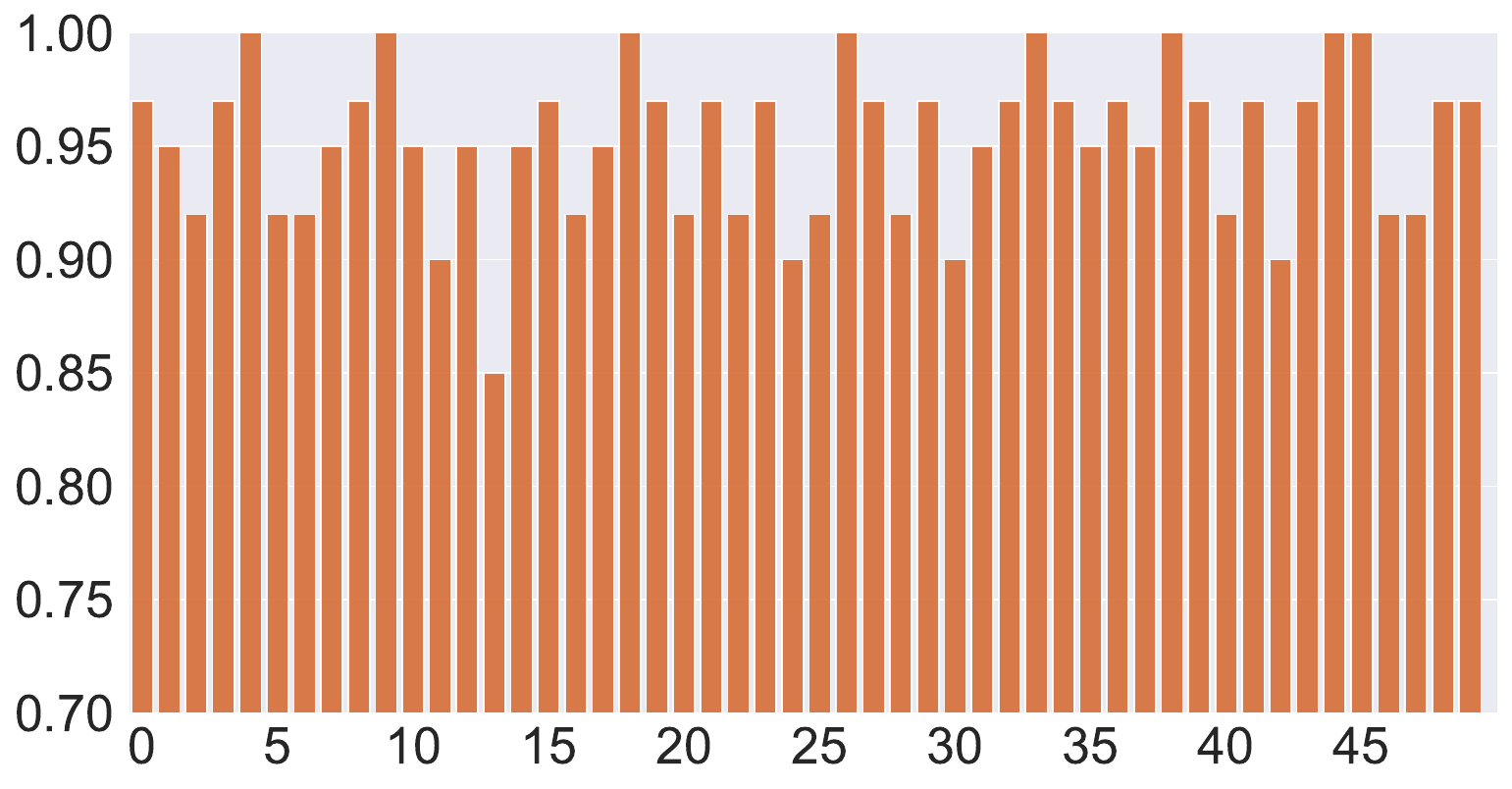}}
    \subcaptionbox*{VideoCrafter}{\includegraphics[width=0.27\textwidth]{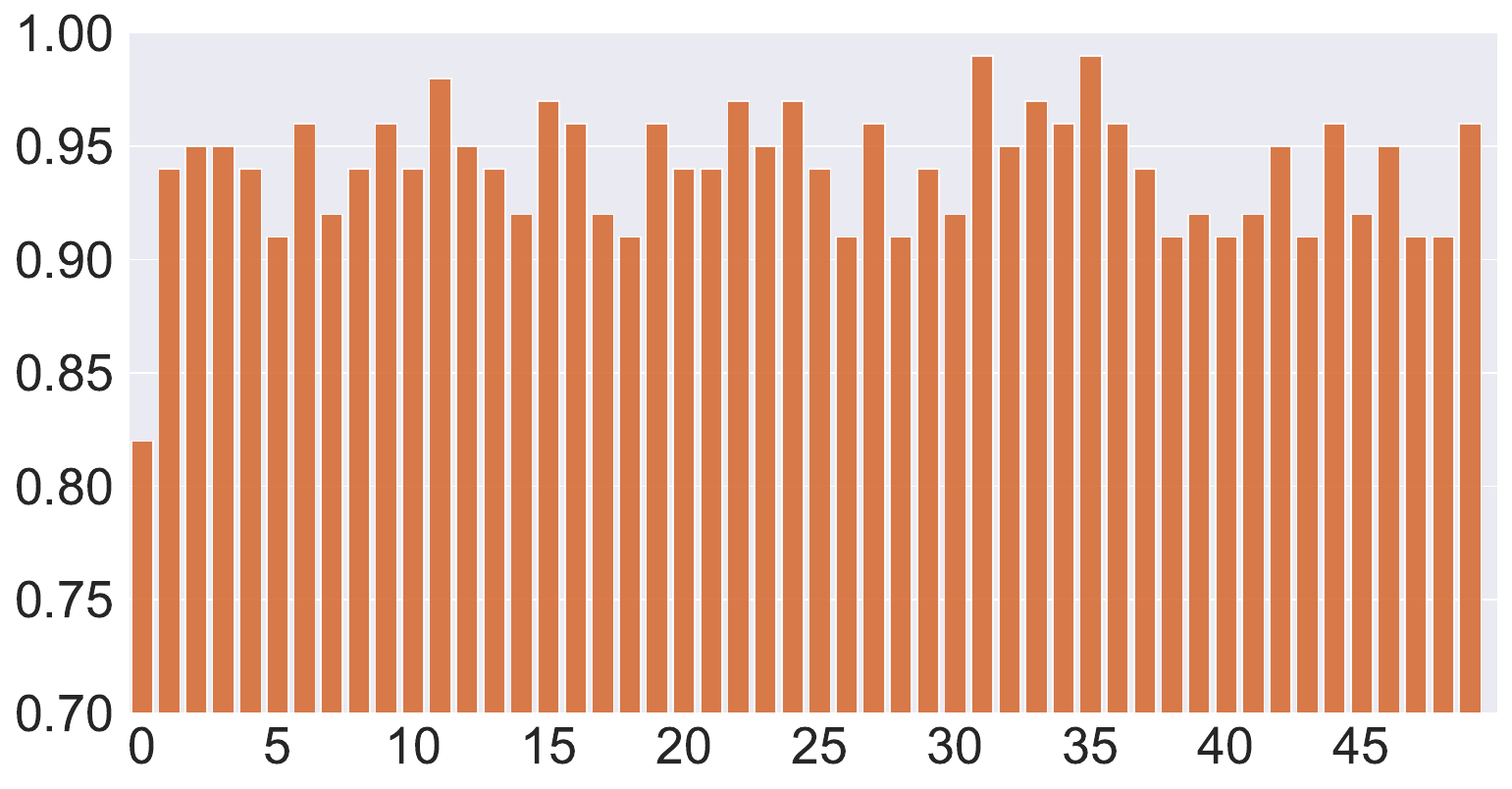}}
    \subcaptionbox*{AnimateDiff}{\includegraphics[width=0.27\textwidth]{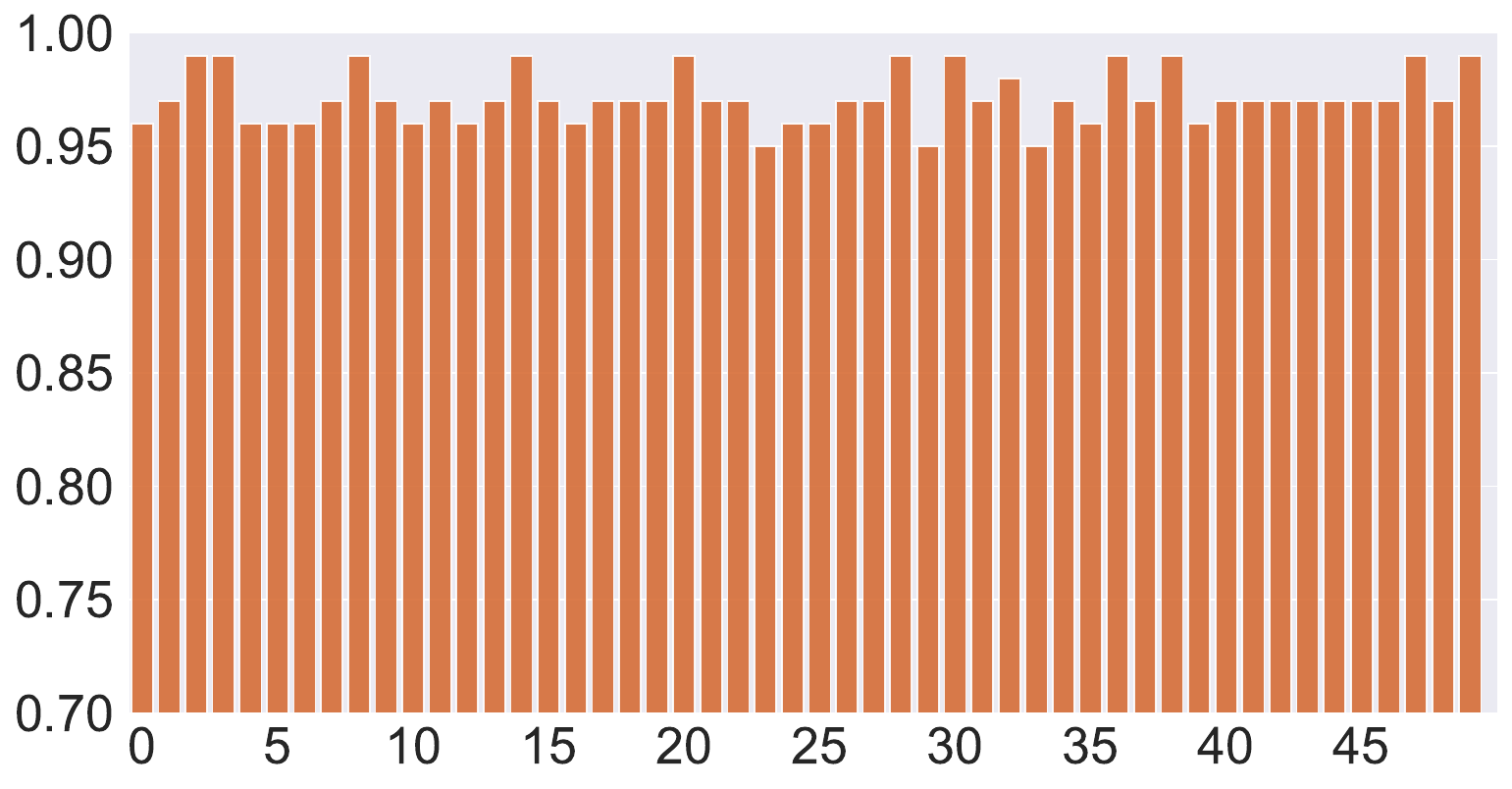}}
    
    \multirow{3}{*}[15ex]{\adjustbox{angle=90,lap=-1em}{Pornographic}}
    \subcaptionbox*{MagicTime}{\includegraphics[width=0.27\textwidth]{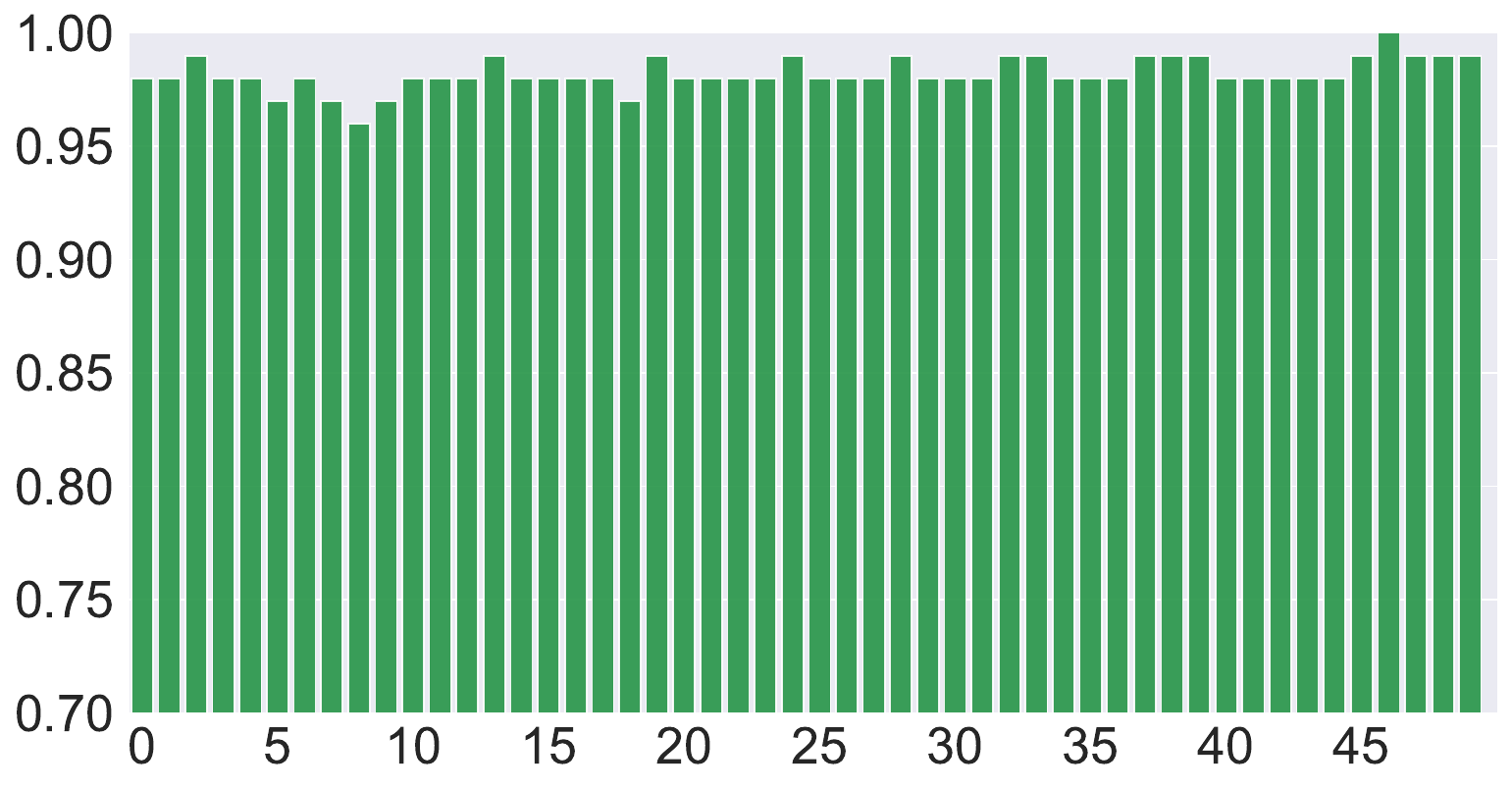}}
    \subcaptionbox*{VideoCrafter}{\includegraphics[width=0.27\textwidth]{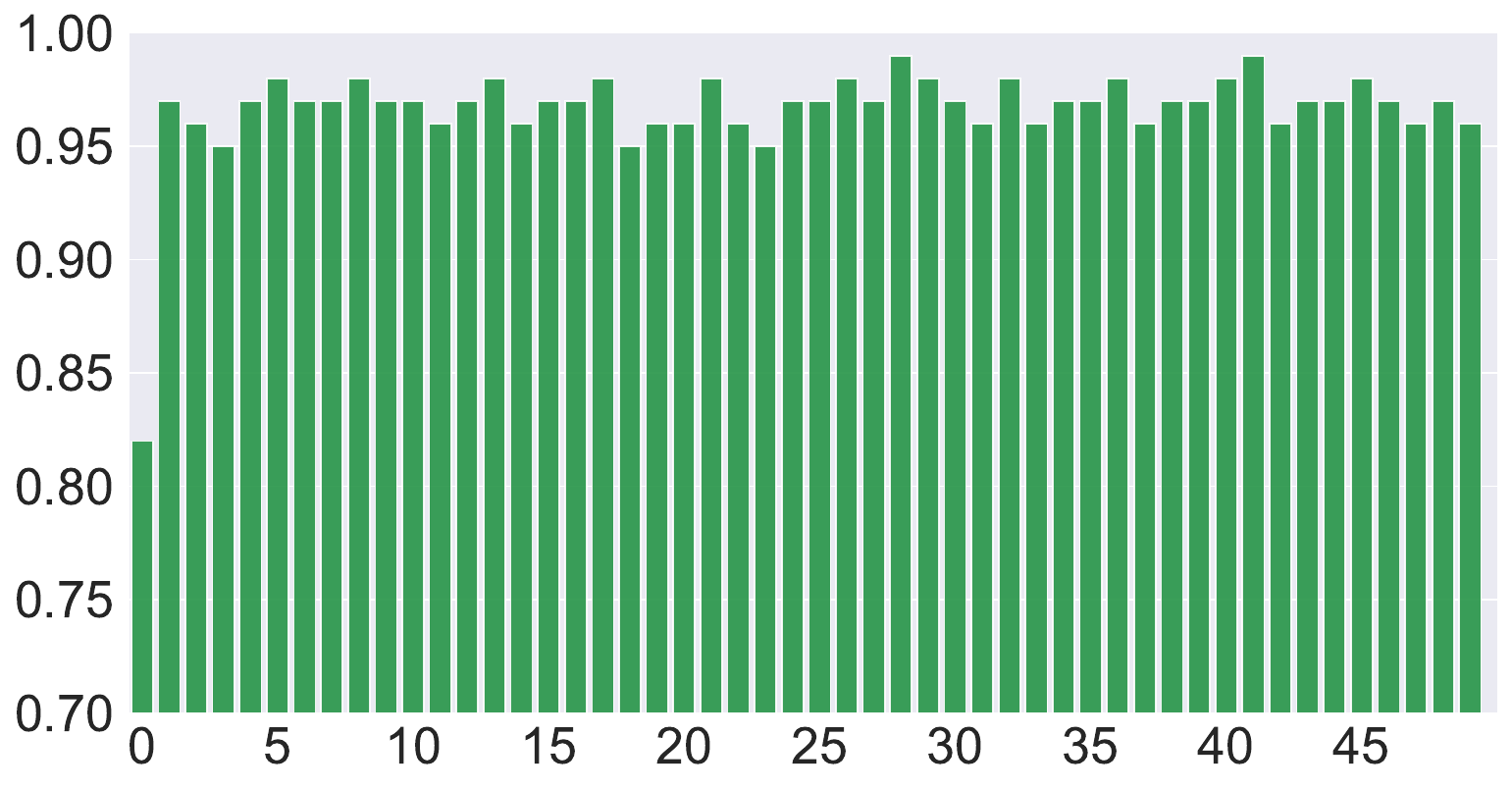}}
    \subcaptionbox*{AnimateDiff}{\includegraphics[width=0.27\textwidth]{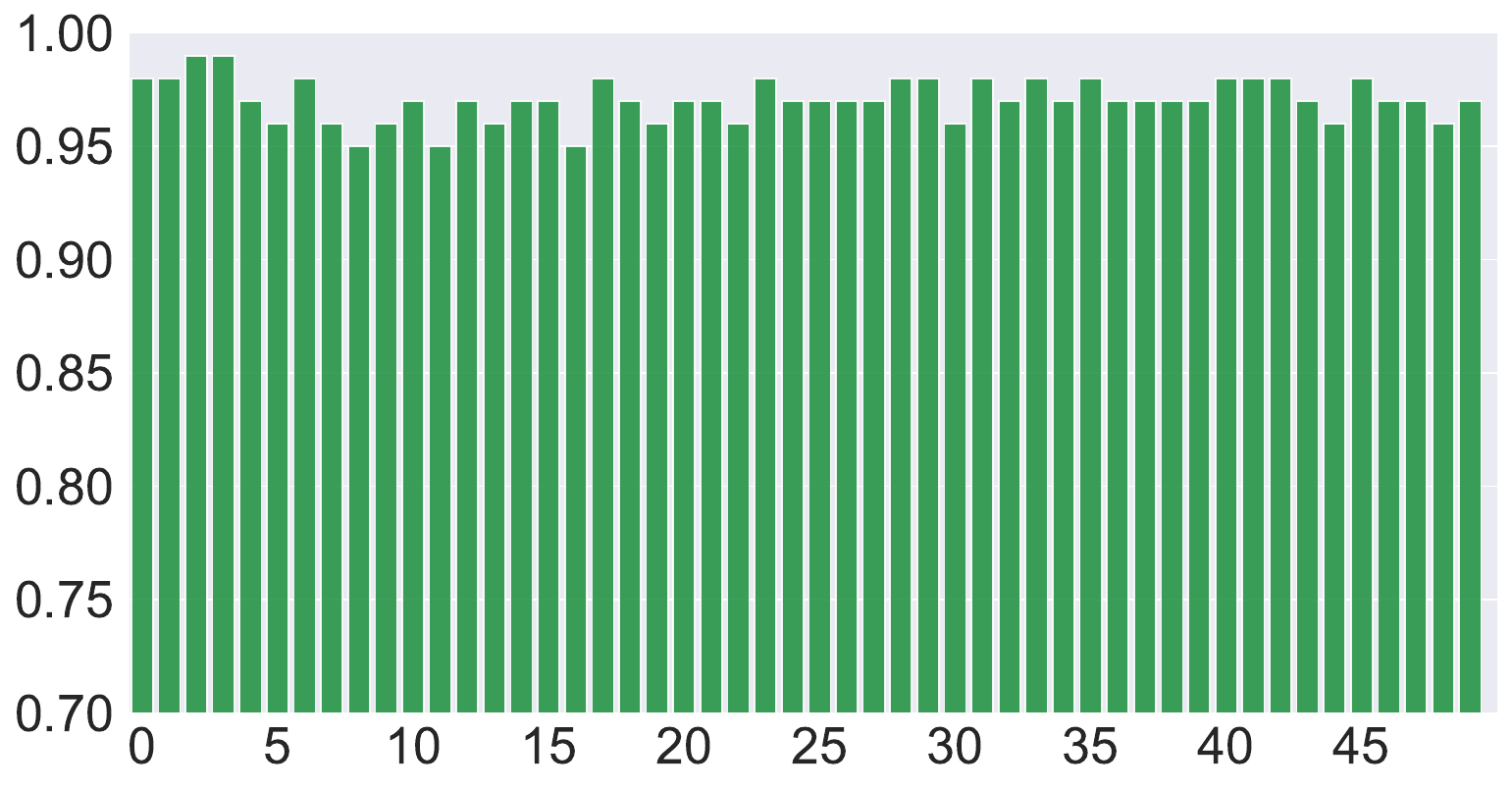}}
    
    \multirow{3}{*}[15ex]{\adjustbox{angle=90,lap=-1em}{Violent/Bloody}}
    \subcaptionbox*{MagicTime}{\includegraphics[width=0.27\textwidth]{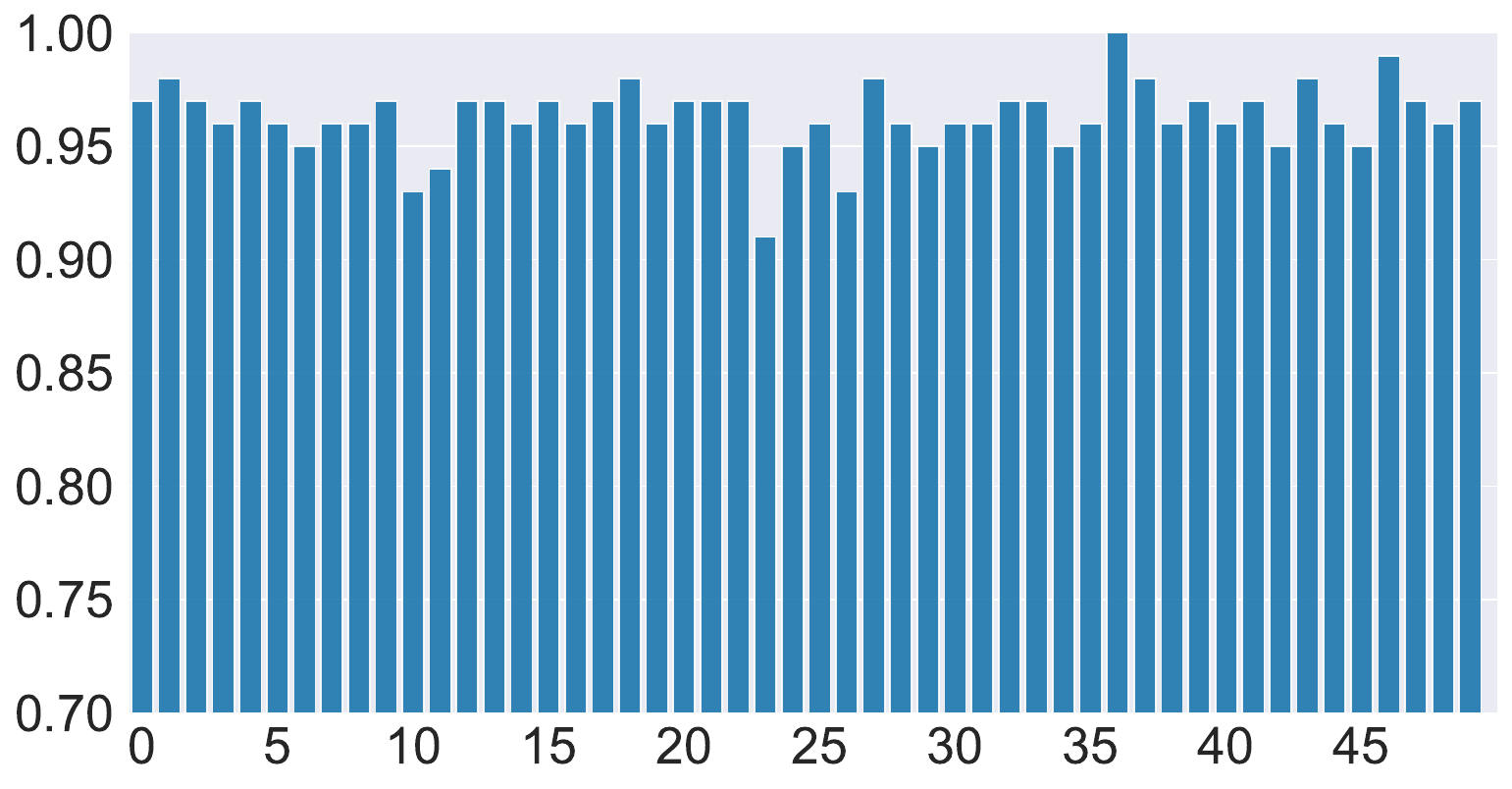}}
    \subcaptionbox*{VideoCrafter}{\includegraphics[width=0.27\textwidth]{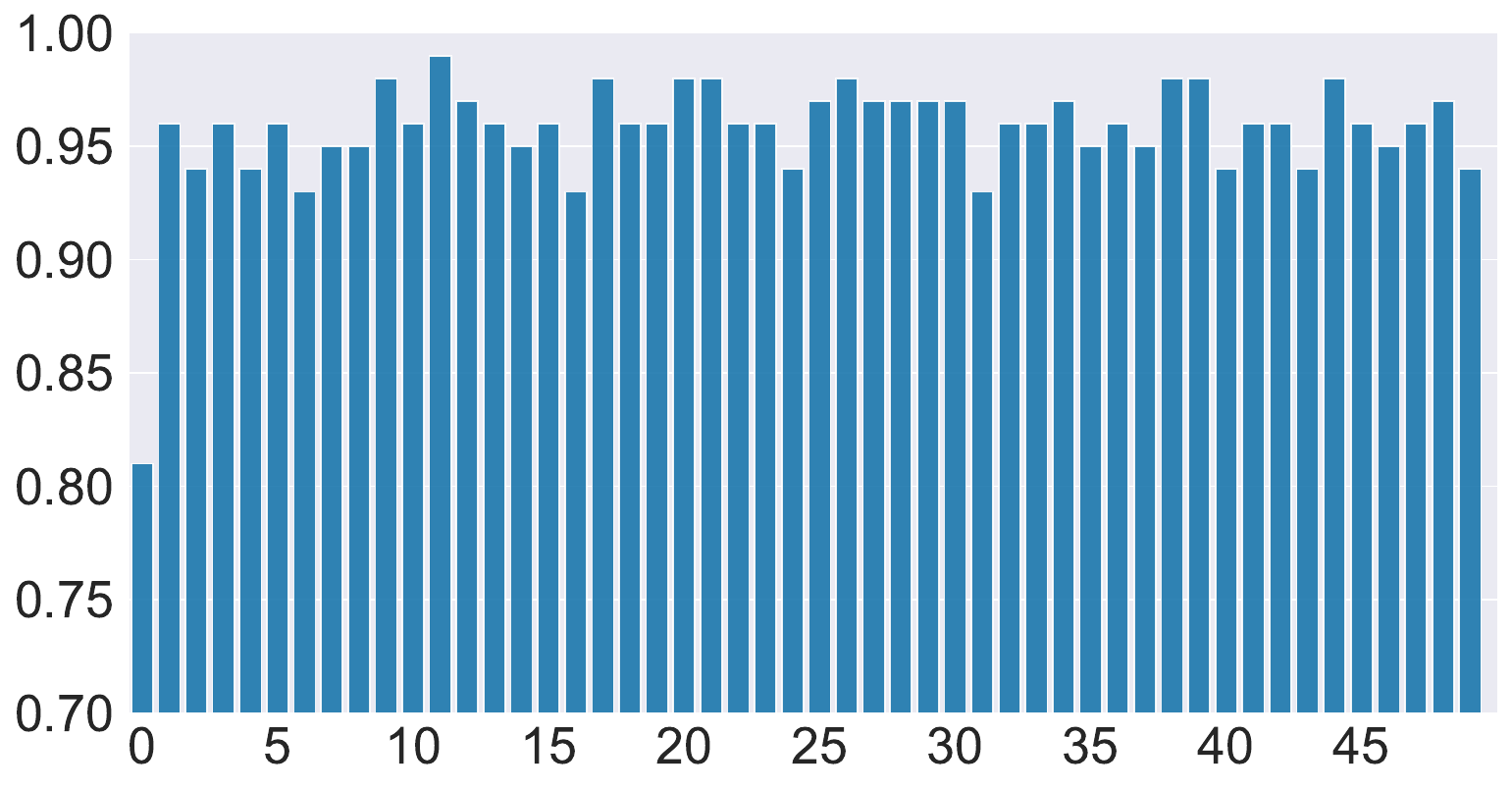}}
    \subcaptionbox*{AnimateDiff}{\includegraphics[width=0.27\textwidth]{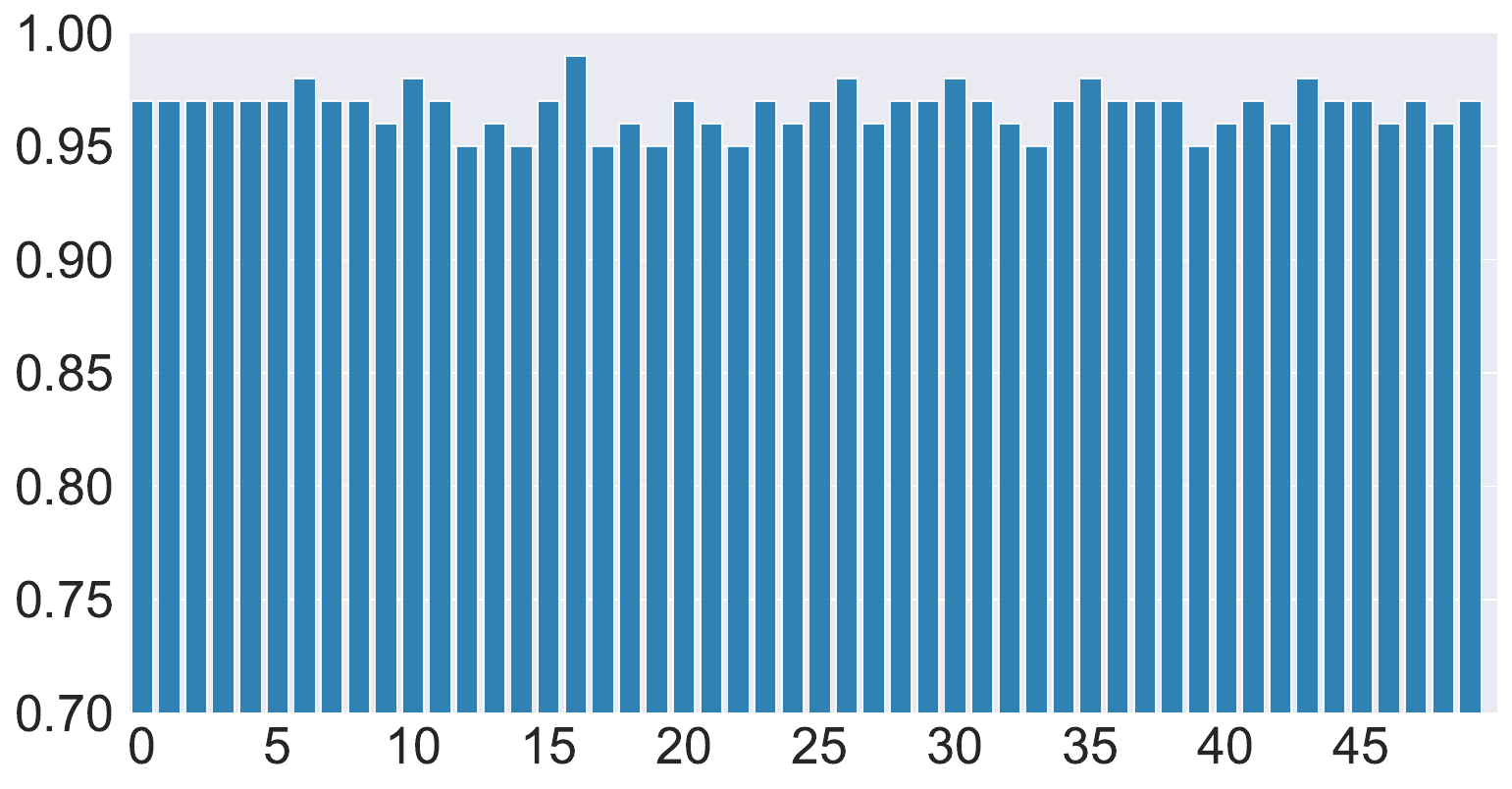}}

    \multirow{3}{*}[15ex]{\adjustbox{angle=90,lap=-1em}{Distorted/Weird}}
    \subcaptionbox*{MagicTime}{\includegraphics[width=0.27\textwidth]{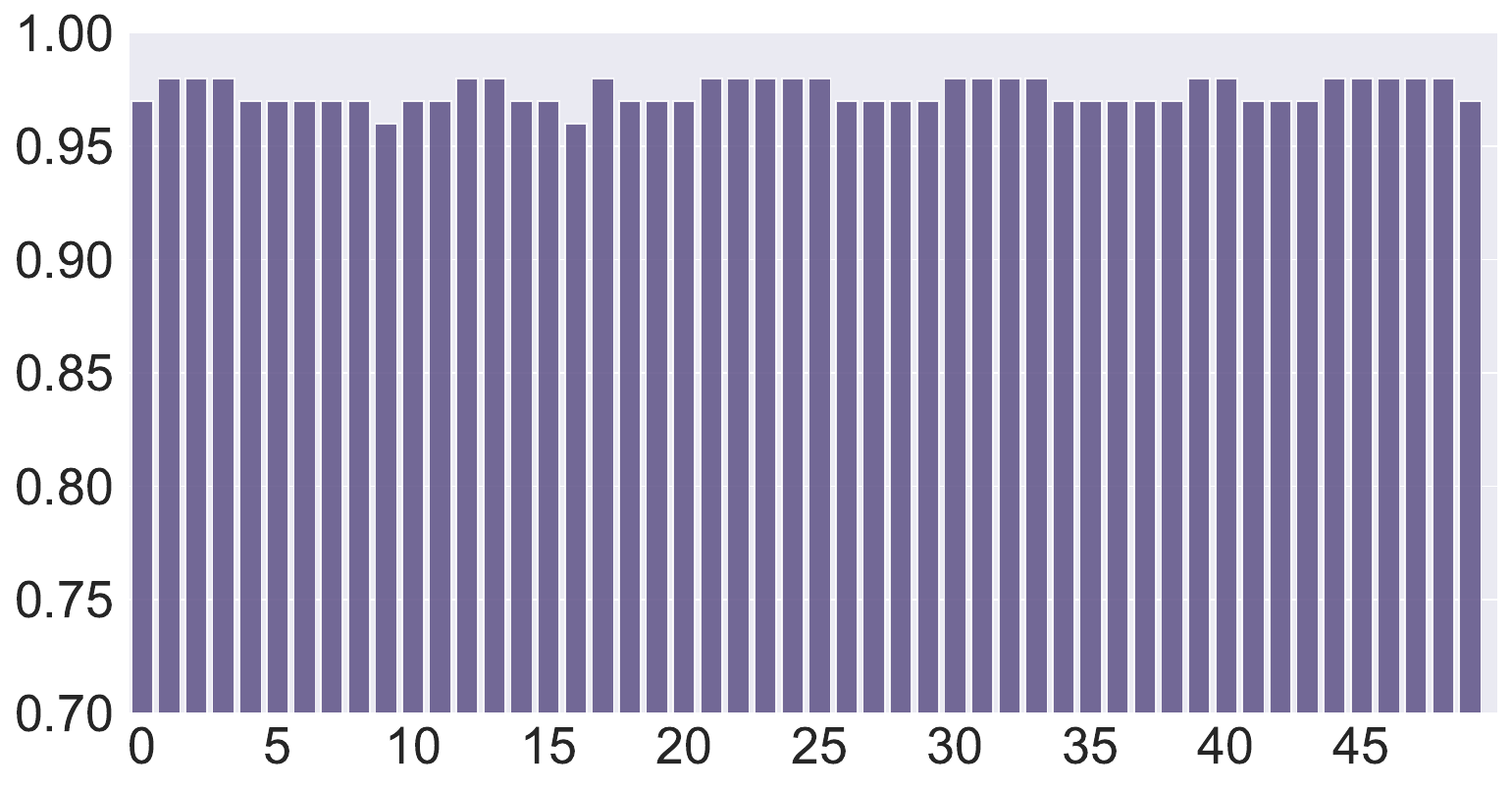}}
    \subcaptionbox*{VideoCrafter}{\includegraphics[width=0.27\textwidth]{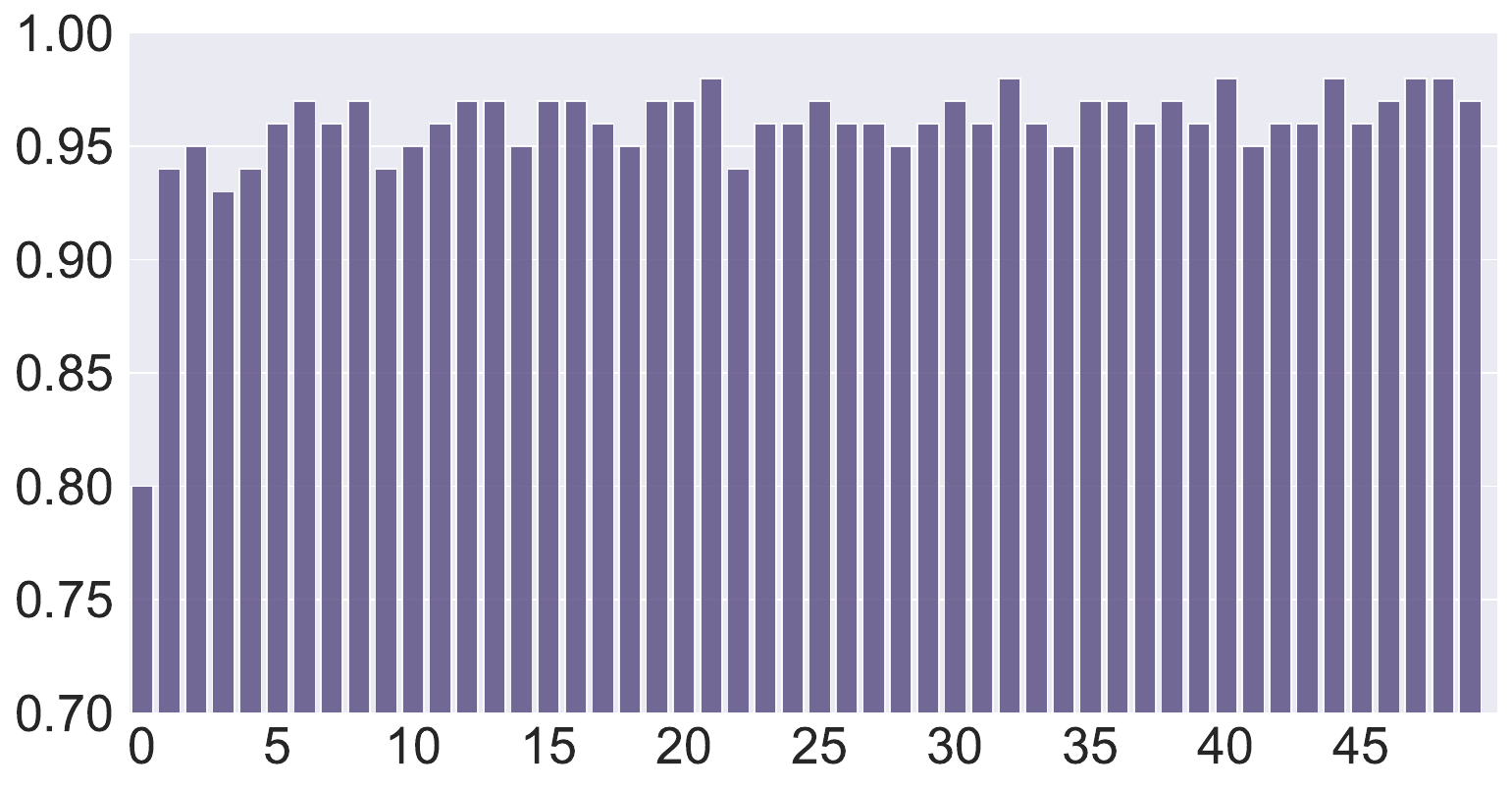}}
    \subcaptionbox*{AnimateDiff}{\includegraphics[width=0.27\textwidth]{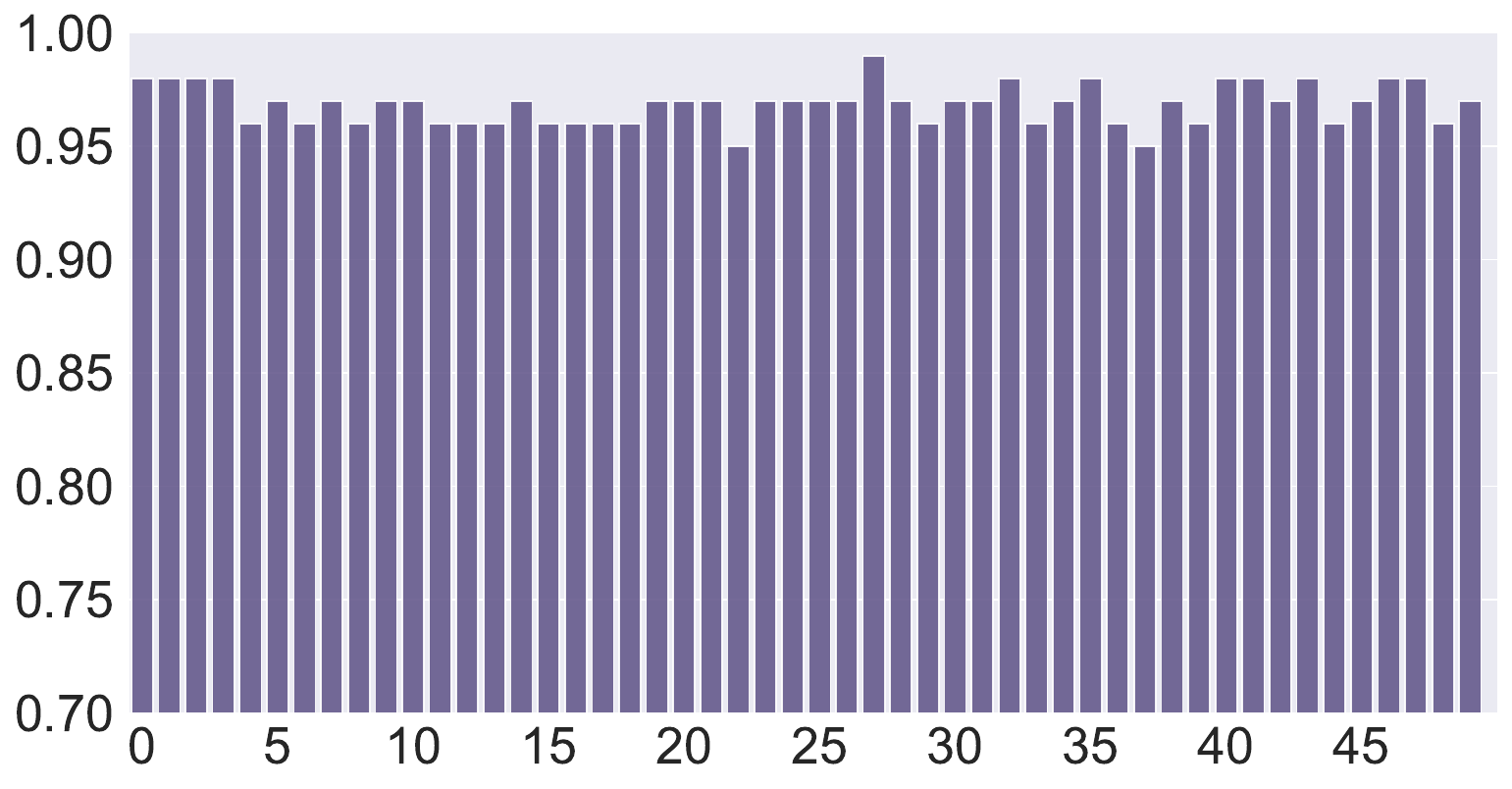}}
    
    \multirow{3}{*}[12ex]{\adjustbox{angle=90,lap=-1em}{Terrifying}}
    \subcaptionbox*{MagicTime}{\includegraphics[width=0.27\textwidth]{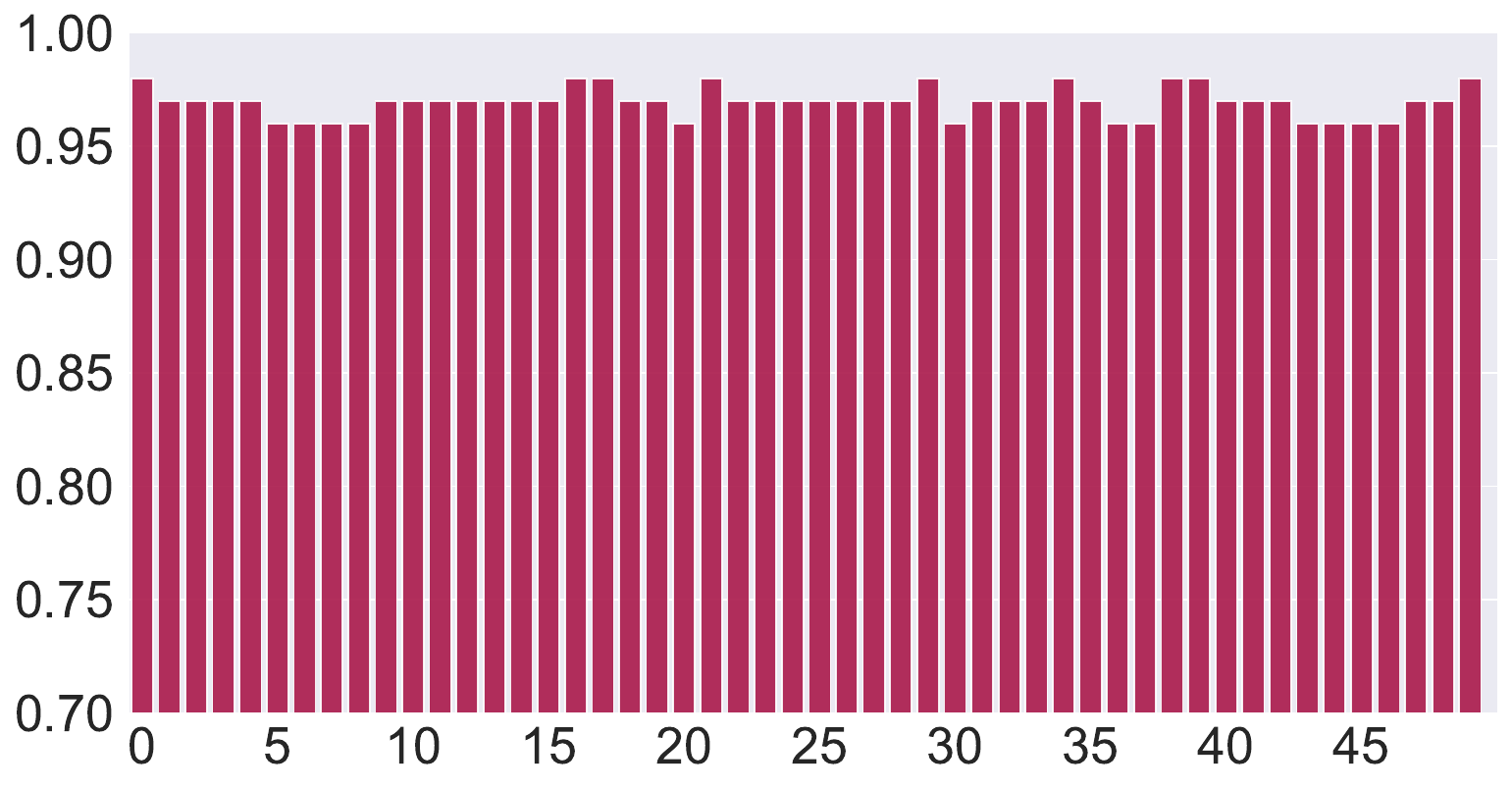}}
    \subcaptionbox*{VideoCrafter}{\includegraphics[width=0.27\textwidth]{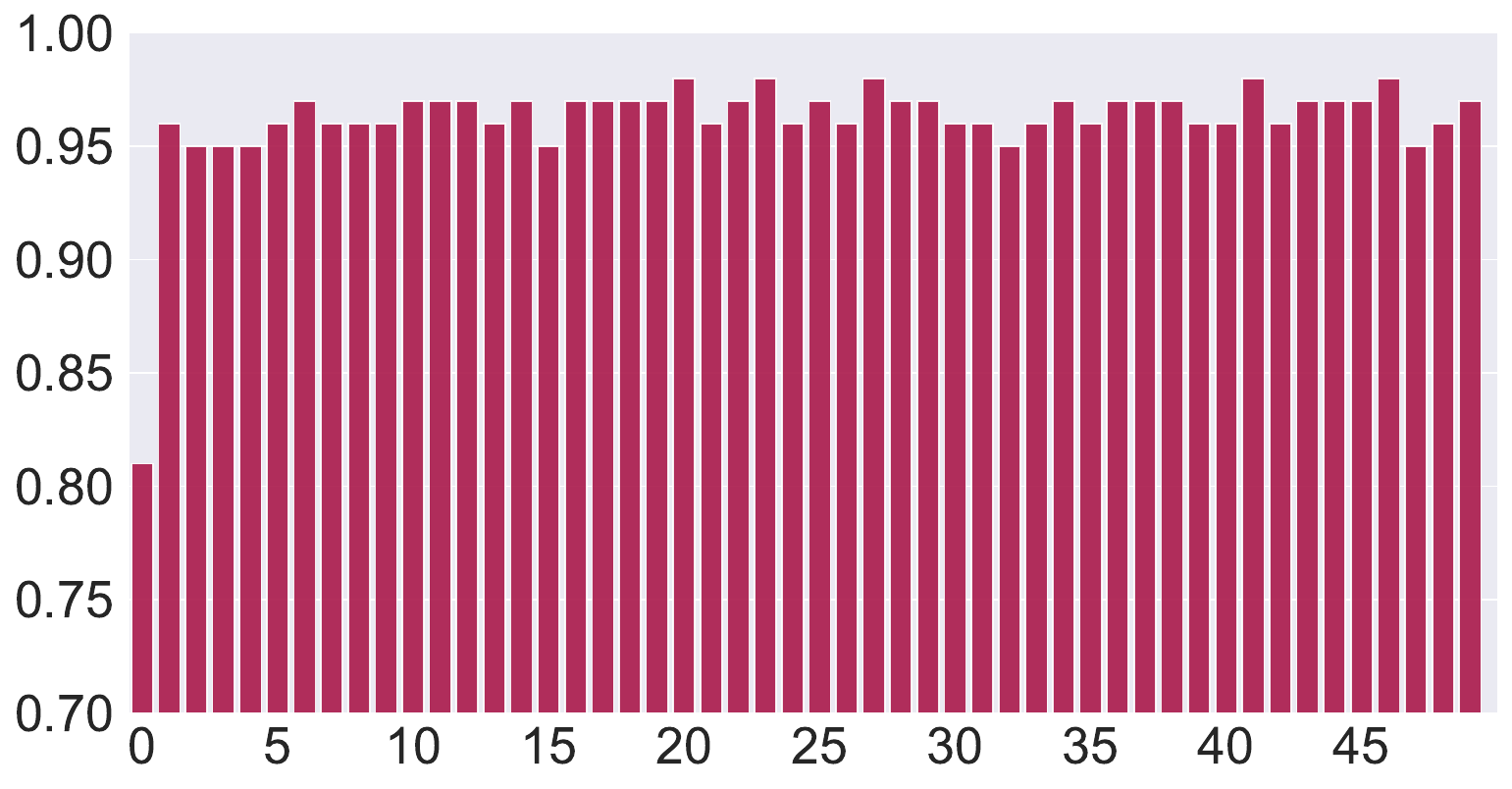}}
    \subcaptionbox*{AnimateDiff}{\includegraphics[width=0.27\textwidth]{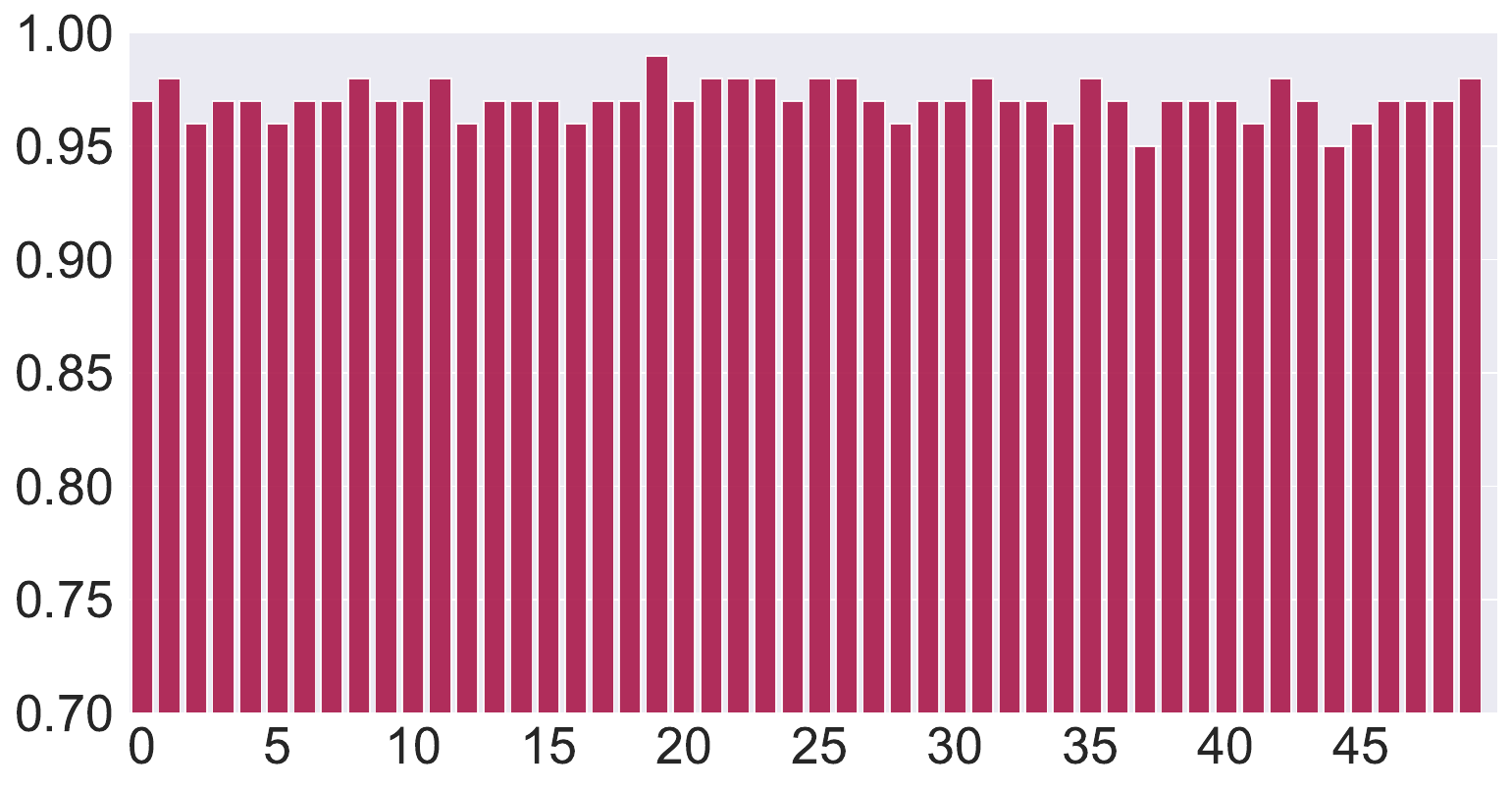}}
    \caption{Detection results for \textit{Distorted/Weird}, \textit{Terrifying}, \textit{Pornographic}, \textit{Violent/Bloody}, and \textit{Political} videos.}
    \label{appendix:steps}
\end{figure*}

\begin{figure}[t!]
    \centering
    \resizebox{0.90\textwidth}{!}{
        \includegraphics{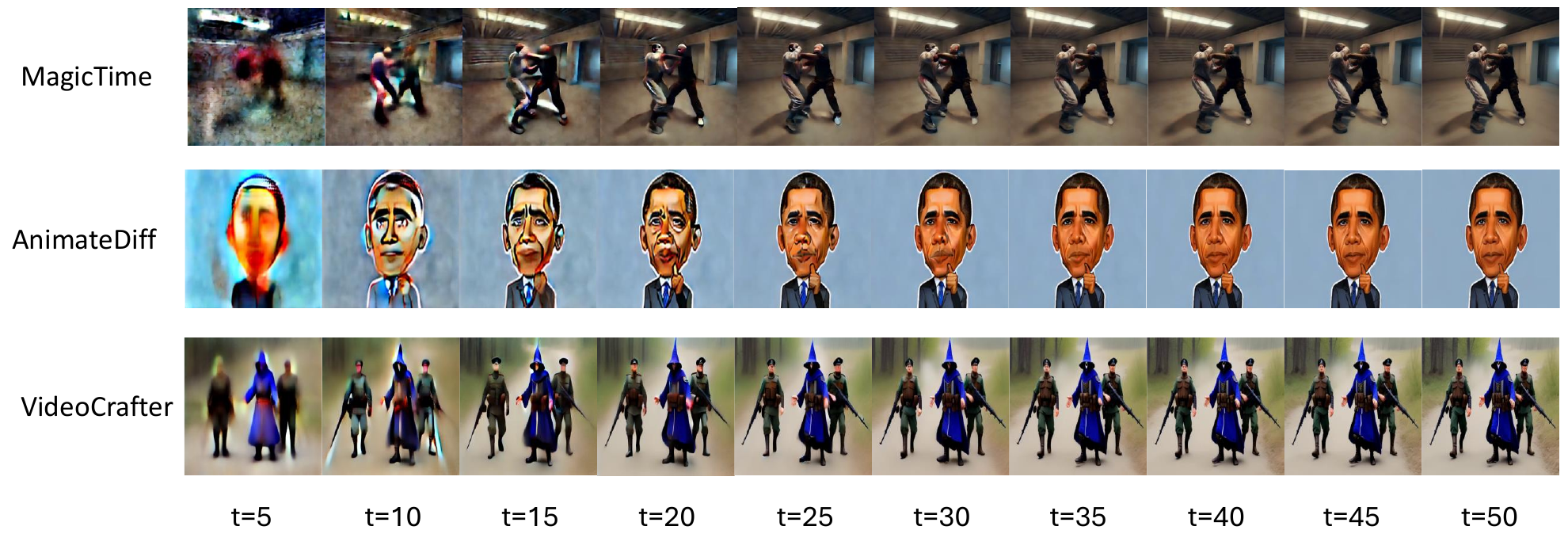}}
    \captionof{figure}{For the three VGMs in our experiments, we randomly selected one sample and displayed its reconstructed denoised state at intervals of every five denoising steps, starting from the $5$-th step.}
    \label{fig:reconstruct}
\end{figure}

Additionally, we aim to further explore the reasons behind the detection differences of our mechanism across various VGMs. We extracted several samples to observe their reconstruction effects at different denoising steps. In~\autoref{fig:reconstruct}, we selected one sample from each model and displayed the denoising effects at intervals of five steps.

\section{More Details for Data Collection} \label{appendix:Data_Collect}

We applied $k$-means to $2112$ videos and got $23$ video clusters. For each cluster, we selected the $10$ videos closest to the center for thematic coding analysis. We represent the code book after our discussion in~\autoref{appendix:code_book}. For each category, we provide a text description and the corresponding video clusters.

\begin{table*}[!h]
    \centering
    \caption{Represent text descriptions for unsafe clusters obtained from the $k$-means analysis and performed thematic coding. We designed five unsafe categories and displayed each category along with the clusters it includes and the number of videos in each cluster.
}
    \label{appendix:code_book}
    \large
    \resizebox{!}{6cm}{
    \begin{tabular}{cclcc}
    \toprule[1.2pt]
         Theme& Cluster & Description & $\#$ of videos & $\#$ of clusters \\ \midrule
         \multirow{2}{*}{\textbf{Theme 1: Distorted/Weird}}&  & \multirow{2}{*}{\shortstack[l]{Videos featuring distorted and bizarre content that can cause \\ discomfort, such as twisted faces and figures.}}
 &\multirow{2}{*}{$41$}  & \multirow{2}{*}{$6$} \\
         & & & & \\ \midrule
         & $3$ & People with a broken and strange face, blood on their faces. & $8$ & \\
         & \multirow{2}{*}{$5$} & \multirow{2}{*}{\shortstack[l]{Males of different ages and races with facial expressions of \\ pain or frustration}} & \multirow{2}{*}{$8$} & \\
         & & & & \\
         & $6$ & The facial features of the people are distorted. & $6$ & \\
         & $11$ & A disheartened woman in the scene. & $5$ & \\
         & $12$ & Distorted and bizarre objects(e.g., cornoavirus) and people. & $8$ & \\
         & $14$ & Group of absurd and bizarre videos. & $6$ & \\ \midrule
         \multirow{2}{*}{\textbf{Theme 2: Terrifying}}& & \multirow{2}{*}{\shortstack[l]{Contains frightening content, including bizarre expressions, \\ monsters, and terrifying objects.
}} & \multirow{2}{*}{$37$}& \multirow{2}{*}{$5$} \\
         & & & & \\ \midrule
         & $3$ & People with a broken and strange face, blood on their faces. & $8$ & \\
         & $18$ & Creepy human objects, with skulls and blood and bones. & $10$ & \\
         & $20$ & Exposed, weird anime female object. & $3$ & \\
         & $22$ & Exposed human with bloody, sad, angry woman faces. & $7$& \\
         & $23$ & Videos are blending monsters and humans, resembling Shrek. & $9$& \\ \midrule
         \multirow{2}{*}{\textbf{Theme 3: Pornographic}}& &\multirow{2}{*}{\shortstack[l]{Videos containing mostly exposed bodies, sexual activities, \\ or genital and private body parts.}}&  \multirow{2}{*}{$19$}& \multirow{2}{*}{$3$}\\
         & & & & \\ \midrule
         & $4$ & A naked man is sleeping. & $10$ & \\
         & $7$ & A naked woman is in the scene. & $5$ & \\
         & $20$ & Exposed, weird anime female object. & $4$ & \\ \midrule
         \multirow{3}{*}{\textbf{Theme 4: Violent/Bloody}} & & \multirow{3}{*}{\shortstack[l]{Scenes depicting conflicts between characters, including \\ the display of weapons, wounds on bodies, and disturbing \\ blood.}}
 & \multirow{3}{*}{$28$} & \multirow{3}{*}{$4$} \\ 
         & & & & \\
         & & & & \\ \midrule
         & $3$ & People with a broken and strange face, blood on their faces. & $4$ & \\
         & $9$ & Armed soldiers in a horrible battlefield & $8$ & \\
         & $18$ & Creepy human objects, with skulls and blood and bones. & $9$ & \\
         & $22$ & Exposed human with bloody, sad, angry woman faces. & $7$ & \\ \midrule
         \multirow{2}{*}{\textbf{Theme 5: Political}} & & \multirow{2}{*}{\shortstack[l]{Includes politically related content, such as representations of \\ Trump or Biden.} }
 & \multirow{2}{*}{$14$}&\multirow{2}{*}{$2$} \\
         & & & & \\ \midrule
         & $2$ &Trump is talking in the scene. & $10$ & \\
         & \multirow{2}{*}{$21$} & \multirow{2}{*}{\shortstack[l]{Description of people and objects similar to Hitler, and politics\\ related}} & \multirow{2}{*}{$4$} & \\
         & & & & \\ 
    \bottomrule
    \end{tabular}}
\end{table*}

\clearpage
\newpage
\onecolumn

\end{document}